\documentclass[journal,transmag]{IEEEtran}
\usepackage{times}
\usepackage{epsfig}
\usepackage{graphicx}
\usepackage{amsmath}
\usepackage{amssymb}
\usepackage{url}
\usepackage{cite}
\usepackage{booktabs}
\usepackage{multirow}
\usepackage{siunitx}
\usepackage{listings}
\usepackage[pagebackref=true,breaklinks=true,colorlinks,bookmarks=false]{hyperref}

\usepackage{cite}
\ifCLASSINFOpdf
\else
\fi
\usepackage{amsmath}
\usepackage{algorithmic}
\usepackage{array}
\usepackage{url}
\usepackage{listings, xcolor}

\definecolor{verylightgray}{rgb}{.97,.97,.97}

\lstdefinelanguage{Solidity}{
	keywords=[1]{anonymous, assembly, assert, balance, break, call, callcode, case, catch, class, constant, continue, constructor, contract, debugger, default, delegatecall, delete, do, else, emit, event, experimental, export, external, false, finally, for, function, gas, if, implements, import, in, indexed, instanceof, interface, internal, is, length, library, log0, log1, log2, log3, log4, memory, modifier, new, payable, pragma, private, protected, public, pure, push, require, return, returns, revert, selfdestruct, send, solidity, storage, struct, suicide, super, switch, then, this, throw, transfer, true, try, typeof, using, value, view, while, with, addmod, ecrecover, keccak256, mulmod, ripemd160, sha256, sha3}, % generic keywords including crypto operations
	keywordstyle=[1]\color{blue}\bfseries,
	keywords=[2]{address, bool, byte, bytes, bytes1, bytes2, bytes3, bytes4, bytes5, bytes6, bytes7, bytes8, bytes9, bytes10, bytes11, bytes12, bytes13, bytes14, bytes15, bytes16, bytes17, bytes18, bytes19, bytes20, bytes21, bytes22, bytes23, bytes24, bytes25, bytes26, bytes27, bytes28, bytes29, bytes30, bytes31, bytes32, enum, int, int8, int16, int24, int32, int40, int48, int56, int64, int72, int80, int88, int96, int104, int112, int120, int128, int136, int144, int152, int160, int168, int176, int184, int192, int200, int208, int216, int224, int232, int240, int248, int256, mapping, string, uint, uint8, uint16, uint24, uint32, uint40, uint48, uint56, uint64, uint72, uint80, uint88, uint96, uint104, uint112, uint120, uint128, uint136, uint144, uint152, uint160, uint168, uint176, uint184, uint192, uint200, uint208, uint216, uint224, uint232, uint240, uint248, uint256, var, void, ether, finney, szabo, wei, days, hours, minutes, seconds, weeks, years},	% types; money and time units
	keywordstyle=[2]\color{teal}\bfseries,
	keywords=[3]{block, blockhash, coinbase, difficulty, gaslimit, number, timestamp, msg, data, gas, sender, sig, value, now, tx, gasprice, origin},	% environment variables
	keywordstyle=[3]\color{violet}\bfseries,
	identifierstyle=\color{black},
	sensitive=false,
	comment=[l]{//},
	morecomment=[s]{/*}{*/},
	commentstyle=\color{gray}\ttfamily,
	stringstyle=\color{red}\ttfamily,
	morestring=[b]',
	morestring=[b]"
}

\begin{document}
%
% paper title
% Titles are generally capitalized except for words such as a, an, and, as,
% at, but, by, for, in, nor, of, on, or, the, to and up, which are usually
% not capitalized unless they are the first or last word of the title.
% Linebreaks \\ can be used within to get better formatting as desired.
% Do not put math or special symbols in the title.
\title{DLBC: A Deep Learning-Based Consensus in Blockchains for Deep Learning Services}

% author names and affiliations
% transmag papers use the long conference author name format.

\author{\IEEEauthorblockN{
Boyang Li,~\IEEEmembership{Member,~IEEE}, 
Changhao Chenli,
Xiaowei Xu,~\IEEEmembership{Member,~IEEE},\\
Yiyu Shi,~\IEEEmembership{Senior Member,~IEEE}, and
Taeho Jung,~\IEEEmembership{Member,~IEEE}
}

% <-this % stops an unwanted space
\thanks{The authors are with the Department of Computer Science and Engineering, University of Notre Dame, Notre Dame, IN 46556 USA. E-mail: \{bli1, cchenli, xxu8, yshi4, tjung\}@nd.edu).}
}

% The paper headers
\markboth{IEEE Transactions on Services Computing,~Vol.~xx, No.~xx, MONTH~20XX}%
{Shell \MakeLowercase{\textit{et al.}}: Bare Demo of IEEEtran.cls for IEEE Transactions on Services Computing}
% The only time the second header will appear is for the odd numbered pages
% after the title page when using the twoside option.
% 
% *** Note that you probably will NOT want to include the author's ***
% *** name in the headers of peer review papers.                   ***
% You can use \ifCLASSOPTIONpeerreview for conditional compilation here if
% you desire.

% If you want to put a publisher's ID mark on the page you can do it like
% this:
%\IEEEpubid{0000--0000/00\$00.00~\copyright~2015 IEEE}
% Remember, if you use this you must call \IEEEpubidadjcol in the second
% column for its text to clear the IEEEpubid mark.

% use for special paper notices
%\IEEEspecialpapernotice{(Invited Paper)}

% for Transactions on Magnetics papers, we must declare the abstract and
% index terms PRIOR to the title within the \IEEEtitleabstractindextext
% IEEEtran command as these need to go into the title area created by
% \maketitle.
% As a general rule, do not put math, special symbols or citations
% in the abstract or keywords.
\IEEEtitleabstractindextext{%
\begin{abstract}
With the increasing artificial intelligence application, deep neural network (DNN) has become an emerging task. However, to train a good deep learning model will suffer from enormous computation cost and energy consumption. Recently, blockchain has been widely used, and during its operation, a huge amount of computation resources are wasted for the Proof of Work (PoW) consensus. In this paper, we propose DLBC to exploit the computation power of “miners” for deep learning training as proof of useful work instead of calculating hash values. it distinguishes itself from recent proof of useful work mechanisms by addressing various limitations of them. Specifically, DLBC handles multiple tasks, larger model and training datasets, and introduces a comprehensive ranking mechanism that considers tasks difficulty(e.g., model complexity, network burden, data size, queue length). We also applied DNN-watermark ~\cite{uchida2017embedding} to improve the robustness.

In Section ~\ref{sec:exp}, the average overhead of digital signature is 1.25, 0.001, 0.002 and 0.98 seconds, respectively, and the average overhead of network is 3.77, 3.01, 0.37 and 0.41 seconds, respectively. Embedding a watermark takes 3 epochs and removing a watermark takes 30 epochs. This penalty of removing watermark will prevent attackers from stealing, improving, and resubmitting DL models from honest miners. 

% add experiment results related to Ownership-protection.

% also, add the results to the introduction.

\end{abstract}

% Note that keywords are not normally used for peerreview papers.
}

% make the title area
\maketitle

% To allow for easy dual compilation without having to reenter the
% abstract/keywords data, the \IEEEtitleabstractindextext text will
% not be used in maketitle, but will appear (i.e., to be "transported")
% here as \IEEEdisplaynontitleabstractindextext when the compsoc 
% or transmag modes are not selected <OR> if conference mode is selected 
% - because all conference papers position the abstract like regular
% papers do.
\IEEEdisplaynontitleabstractindextext
% \IEEEdisplaynontitleabstractindextext has no effect when using
% compsoc or transmag under a non-conference mode.

% For peer review papers, you can put extra information on the cover
% page as needed:
% \ifCLASSOPTIONpeerreview
% \begin{center} \bfseries EDICS Category: 3-BBND \end{center}
% \fi
%
% For peerreview papers, this IEEEtran command inserts a page break and
% creates the second title. It will be ignored for other modes.
\IEEEpeerreviewmaketitle

%------------------------Begin

%%%%%%%%% ABSTRACT

%%%%%%%%% BODY TEXT
\section{Introduction} \label{sec:intro}
\IEEEPARstart{B}{itcoin}~\cite{nakamoto2008bitcoin} is the most popular blockchain technology-based application. Besides countless cryptocurrencies, blockchain technology has been successfully applied in different fields. However, the traditional Proof-of-Work (PoW) consensus mechanism demands an immense amount of energy for computation to maintain the blockchain. According to Digiconomist~\cite{Bitcoin}, %in Fig.~\ref{fig:power}, 
the estimated power consumption of Bitcoin ``mining" reaches around 70 TWh per year during the second half of the year 2018. As a result, there are the concerns and warnings about energy wasting of cryptocurrencies~\cite{de2018bitcoin}, for instance, Camilo Mora published a paper in Nature Climate Change with the title of ``Bitcoin emissions alone could push global warming above 2 centigrade"~\cite{mora2018bitcoin}.

To maintain the consistency of transactions, the traditional Proof of Work (PoW) consensus mechanism utilizes the brute-force algorithms to host a competition of hardware and energy source, and this is the major component that leads to the energy wasting issue. A series of solutions have been proposed to address this issue, such as ASIC machine~\cite{narayanan2016bitcoin}, Proof of Stake (PoS)~\cite{king2012ppcoin}, Proof of Capacity (PoC)~\cite{Burstcoin} and Proof of Useful Work (PoUW)~\cite{narayanan2016bitcoin}. ASIC machines compute hash efficiently, but this type of machine is only able to calculate on a certain type of brute-force algorithms and it is relatively inflexible. PoC significantly wastes disk space instead of electricity. PoS mechanism cannot provide solid security as PoW, because determination of the block creators involves efficient computation only, and it does not consume much energy~\cite{romano2017beyond}.
On contrary, PoUW exploits computation power of ``miners" for useful tasks, therefore the energy consumed by the miners is not wasted.

%-----------------------------------------------------------------------------------

%-----------------------------------------------------------------------------------

\iffalse
\begin{figure}[t]
\begin{center}
 \includegraphics[width=.9\columnwidth]{figs/btc_power.pdf}
\end{center}
   \caption{Estimated power consumption of Bitcoin~\cite{Bitcoin}. }
\label{fig:power}
\end{figure} 
\fi

Deep learning plays a crucial role in many fields, for instance, supporting clinic diagnosis %~\cite{xu2018quantization}.
~\cite{tianchen2019,xu2018quantization,xu2018scaling,xu2017edge,xu2018efficient,liu2019compression,yang2019towards}.
%which has been successfully applied in brain MRI segmentation~\cite{pereira2016brain}, lung CT scans segmentation~\cite{jin2018ct} and Cardiac MRI Segmentation~\cite{avendi2016combined}. 
%Instead of segmenting images manually, these deep learning based algorithms accelerate medical diagnosis in terms of extracting different tissues, organs, pathologies, and biological structures. However, it is extremely challenging due to high variability in medical images, low contrast, and other imaging artifacts~\cite{xu2018quantization}. 
By taking advantage of rapid increase of computing power and machine learning research interests \cite{xu2018mda,xu2018accelerating,xu2018resource,xu2017efficient,xu2018efficientHardware,gong2019real,liu2018multi,shen2018practical,yifan2019multi} 
%\cite{xu2018quantization}
, the performance of deep learning has been improved significantly. 

%segmentation based on deep learning is much less expensive and time-consuming than manual segmentation. 
However, it does require huge computation power to achieve high performance model\cite{xu2018scaling,ding2018universal}. In addition, an accurate model requires machine learning experts to tune it by training and evaluating with different hyper-parameters multiple times. Therefore, a good model comes at the cost of very high computation consumption.

%-----------------------------------------------------------------------------------

\begin{figure}[t]
\begin{center}
 \includegraphics[width=.9\columnwidth]{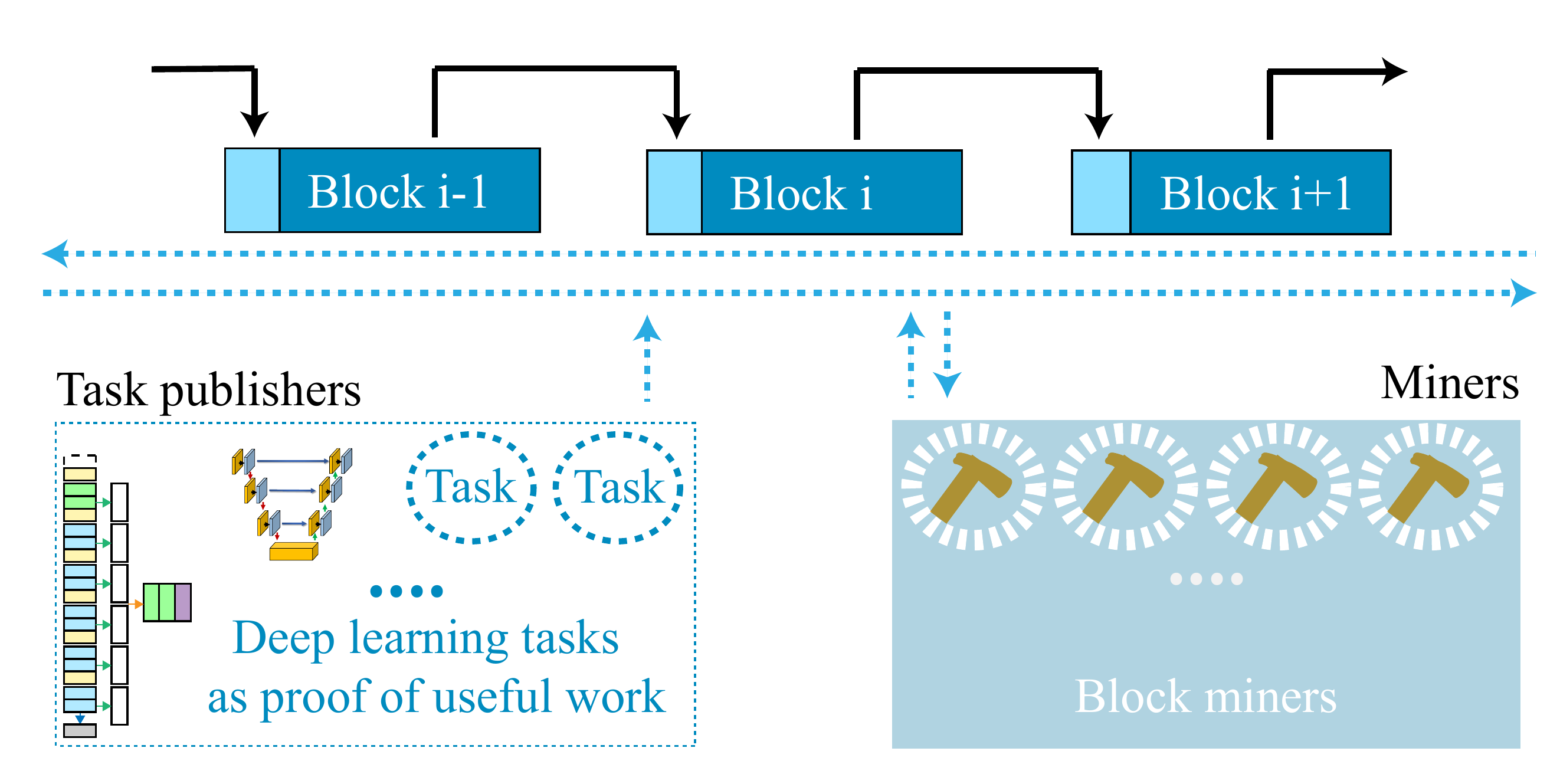}
\end{center}\vspace{-10pt}
   \caption{A blockchain maintained with image Segmentation algorithm as PoUW. }\vspace{-10pt}
\label{fig:teaser}
\end{figure}

Primecoin \cite{king2013primecoin}, PoX \cite{shoker2018brief}, Privacy‐Preserving Blockchain Mining ~\cite{turesson2018deep}, Coin.AI~\cite{baldominos2019coin}, WekaCoin~\cite{bravo2019proof},  and PoDL \cite{chenli2019energy} are PoUW mechanisms that ask miners to perform valuable tasks. 
% alternatives of PoW consensus mechanism that are Po suggest a new type of work rather than hash calculation. Primecoin lets miners find special sequences of prime numbers called Cunningham chain, and PoDL lets miners perform training for deep learning. The Cunningham chain from the Primecoin has limited applications because the prime numbers in the chain are special. The PoDL mechanism demonstrates a protocol that asks miners to train deep learning models and present them as the proof. 
This paper proposes a deep learning based consensus (DLBC) as PoUW, as demonstrated in Fig.~\ref{fig:teaser}, to exploit the computation power of ``miners'' for training deep learning tasks while addressing the limitations of existing PoUW mechanisms. 
This work solved three challenges: (1) Because the ``useful work'' in term of training deep learning model is different from hash algorithm, the original consensus will not fit our case. This novel mechanism has to be fair for all miners which is challenging. (2) Because the DL models and source code are accessible to public, it is necessary and challenging to protect the ownership of all submitted DL models. (3) Because sustainability of this blockchain depends on whether there are enough training tasks, we suggested an mechanism to balance the reward and create incentive for task publisher when the system needs more tasks to train. 

\iffalse
Privacy‐Preserving Blockchain Mining ~\cite{turesson2018deep} proposed two parallel chains solution as a novel consensus. The first chain is for transactions and the second chain for deep learning tasks. 
\fi

In order to address these challenges, we augmented existing memory pool of full nodes to keep all unconfirmed tasks and unselected tasks, a novel task scheduler in each block interval so that miners will work on the same job, and augmented the block data structure to keep the current task and unselected tasks record. %In this design, we also introduced three participants includes the task publisher, miner nodes and full nodes. 
To protect the ownership of the DL models, we leveraged DNN-watermark ~\cite{uchida2017embedding} such that the winner miner can prove and protect the ownership with the embedded watermark in the model. 
Our major contributions are: (1) our mechanism can accept and handle multiple tasks; (2) similar to the state-of-the-art consensus mechanisms based on deep learning, our mechanism can also handle large models and training datasets; (3) we proposed a difficulty score for each submitted task that miners will select a training task based on the scores as the guideline and the mechanism also provide incentive for task publishers; (4) we adopt the DNN-watermark ~\cite{uchida2017embedding} to protect submitted DL models and evaluate the reliability of the embedded watermark method. 

As the overhead evaluation shown in Section~\ref{sec:exp}
indicates, the average overhead of digital signature is 1.25, 0.001, 0.002 and 0.98 seconds, respectively, and the average overhead of network is 3.77, 3.01, 0.37 and 0.41 seconds, respectively. In the watermark evaluation, it demonstrates that embedding a watermark takes 3 epochs and successful removing a watermark takes 30 epochs. The penalty of removing watermark will prevent attackers steeling, improving, and resubmitting DL models from honest miners.

%On top of these existing works, a practical method to exploit the computation power of miners for biomedical image segmentation tasks is proposed in this paper. To achieve better performance, the biomedical image segmentation model will have to contain more parameters that our mechanism is expected to handle bigger models and training data-set. In addition, the task ledger is introduced to handle multiple submission. 

%-----------------------------------------------------------------------------------

%------------------------------------------------------------------------
\section{Background and related work} \label{sec:backg}
%------------------------------------------------------------------------
% !!! will include more deep learning tasks other rather segmentation only. !!!
Multiple machine learning tasks have adopted deep neural networks as solutions and achieved state-of-art results. 

% structure:
% problem definition 
% common network structure and many variations
% lots of layers

\noindent \textbf{Application of deep neural networks in image classification:}
Image classification is a classic computer vision task that has been existing for decades. Classification can be as simple as recognition of handwritten digits such as MNIST dataset. On the other hand, complex classification problems can have thousands of categories involved which sets high accuracy obstacles even for human\cite{grieggs2019measuring}. As deep neural networks triumphed the ImageNet\cite{krizhevsky2012imagenet} challenge since 2012, neural networks have been going deeper and deeper, from AlexNet\cite{ballester2016performance} with 8 layers to as much as ResNet~\cite{szegedy2017inception} with 152 layers. A Major component of these networks is convolutional layers with several filters. As a result, the stack of dozens of convolutional layers usually requires significant amount of computation power.

\begin{figure}[t]
\centering
\vspace{-10pt}
 \includegraphics[width=1\columnwidth]{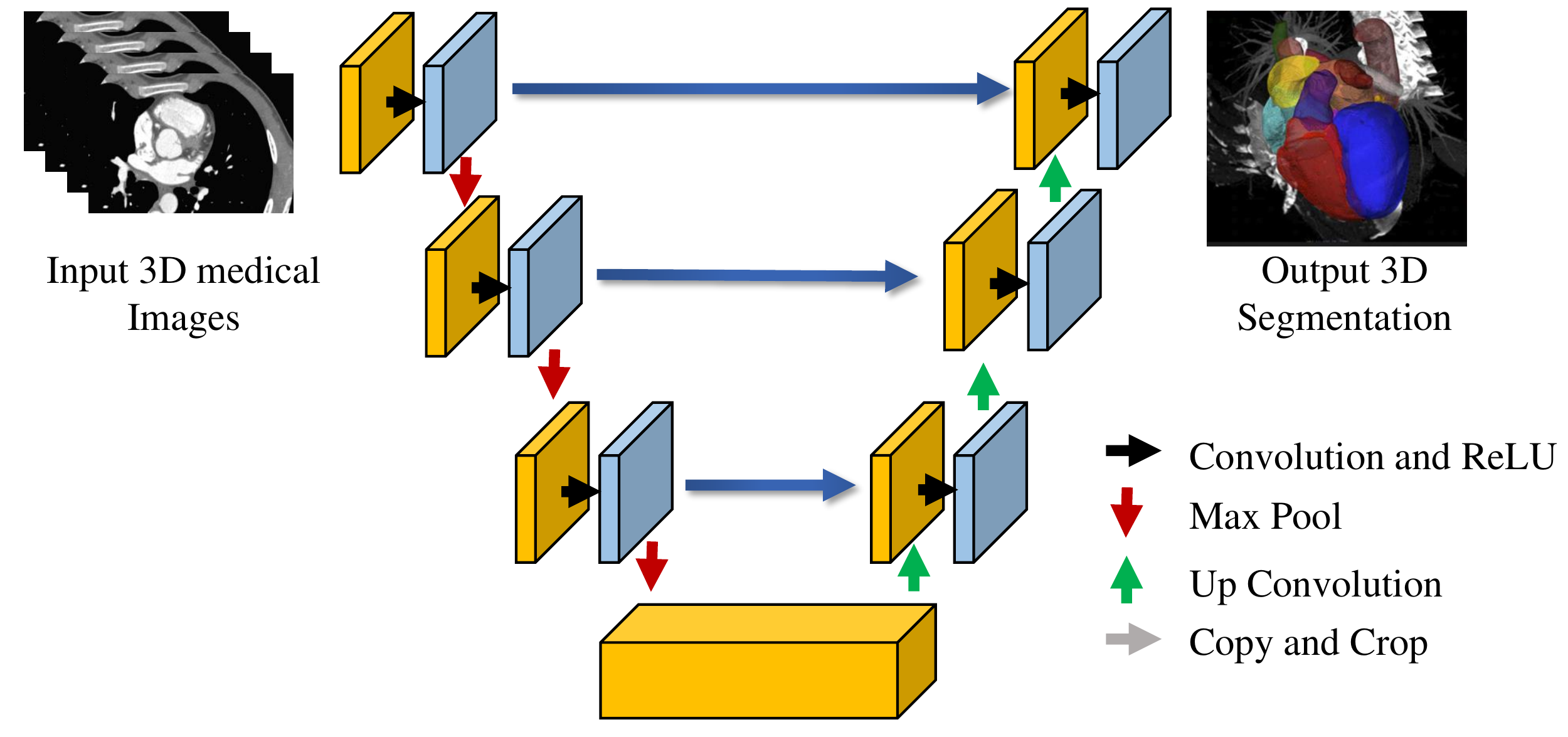}\vspace{-10pt}
 \caption{3D U-Net \cite{dou20163d}: a widely used framework in fully convolutional networks for medical image segmentation.}\vspace{-10pt}
 \label{fcn}
\end{figure}

\noindent \textbf{Application of deep neural networks in biomedical image segmentation:} Biomedical image segmentation is another task which has been boosted tremendously by the development of deep neural networks. The task usually outputs a pixel-wise classification result in the original size of the high-resolution input image. Fully convolutional networks (FCN) is a special category of DNN, which is widely used for medical image segmentation.
Compared with general DNNs, FCNs only consist of convolutional layers, up convolutional layer, and pooling layers as shown in Fig.~\ref{fcn}.
With this characteristic, FCNs can efficiently output images with the same size as the input images as shown in Fig.~\ref{fcn}.
Almost all the DNN-based methods for 3D image segmentation adopt FCN as the backbone network structure, and add some special structures and improve training strategies~\cite{chen2016deep,cciccek20163d,milletari2016v,chen2017voxresnet,dou20163d,wang2019msu,xu2019whole,liu2019machine,xu2017edge}.
For example, 3D U-Net~\cite{cciccek20163d} adds more connections between the first several layers and the last several layers as shown in Fig.~\ref{fcn} to better extract features. With all the details implemented in these convolution operation based deep neural networks and the large image size in this specific problem (square of thousands of pixels for single medical image), training for these models is undoubtedly computationally expensive.

\noindent \textbf{Application of deep neural networks in speech recognition:}
Besides the computer vision area, deep neural networks also dominate the research in speech recognition. Speech recognition aims at translating an input sound wave signal into matching text output. A common approach used to solve the problem would be preprocessing the sound wave into a sequence of slices, then a Recurrent neural network can be applied to generate a corresponding sequence of characters recognized from the input sequence. By training the model with CTC loss, which removes repetition in the output text sequence, the output text can reach a satisfying accuracy.

\begin{figure}[t]
\centering
\vspace{-10pt}
 \includegraphics[width=1\columnwidth]{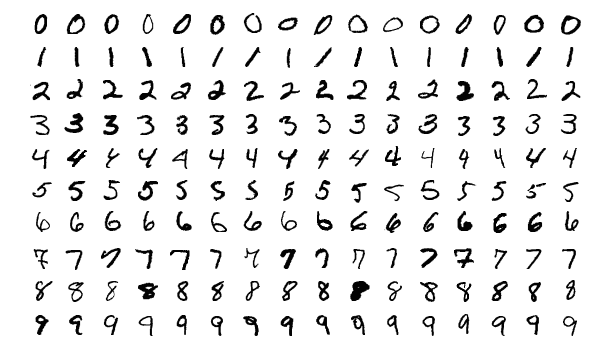}\vspace{-10pt}
 \caption{Hand written digit samples for images classifier tasks (MNIST) ~\cite{lecun1998gradient}}
 \label{mnist}\vspace{-10pt}
\end{figure}

\begin{figure}[t]
\centering
\vspace{-10pt}
 \includegraphics[width=1\columnwidth]{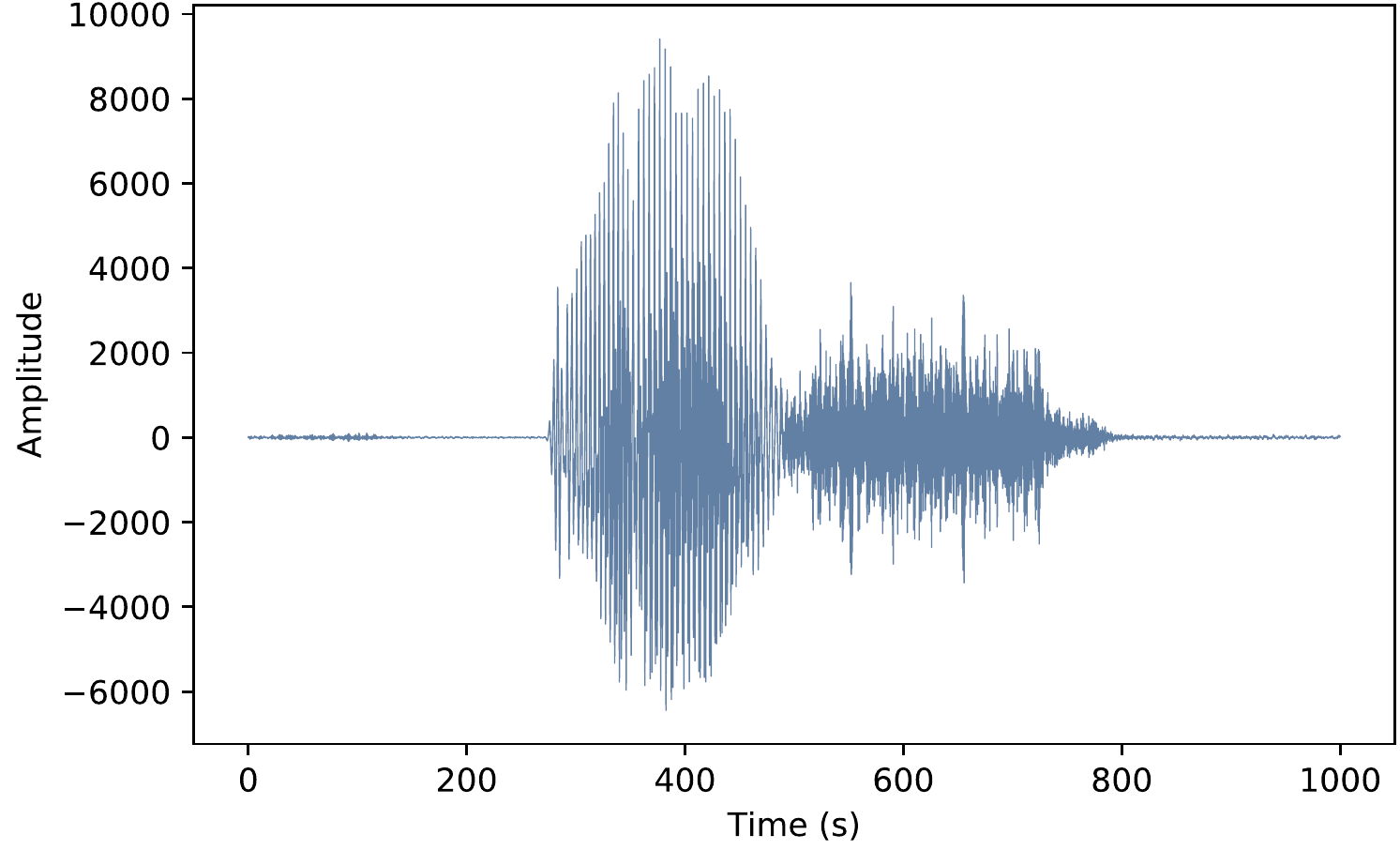}\vspace{-10pt}
 \caption{Single word command audio sample for automatic speech recognition tasks ~\cite{arik2017convolutional}}
 \label{asr}\vspace{-10pt}
\end{figure}

%------------------------------------------------------------------------

%------------------------------------------------------------------------

%------------------------------------------------------------------------
\noindent \textbf{Consensus mechanisms for blockchain:} 
Proof of Stake (PoS) is a consensus mechanism in cryptocurrencies to decide the creator of the next block based on the amount of cryptocurrencies the creator owns or other weights that can prove the authority of the creator. Determination of the block creators involves efficient computation only, and it does not consume much energy. However, it is unknown whether PoS mechanism is a robust distributed consensus mechanism owing to various limitations~\cite{poelstra2014distributed,ogawaproposal,kanjalkarIcant}. Existing cryptocurrencies adopting PoS all have extra rules to make the consensus mechanism more robust, however, the necessity of extra rules implies the PoS is inherently unstable, and the benefit of energy efficiency is diluted.

Proof of Work (PoW) does not suffer from the instability of PoS. Its stability is supported by the enormous computation resources contributed to hard computation problems, and its criticism has been on its large amount of wasted energy. 
The idea of PoW was proposed to prevent from the denial-of-service attacking~\cite{back2002hashcash}.
In PoW consensus, all participants are required to solve problems before they send messages. 
Such problems are challenging to solve and easy to validate. 
In Bitcoin, miner nodes are required to calculate hash values as PoW.

\iffalse
it was firstly proposed in the field of economics to prevent the denial-of-service attack. The basic idea of PoW was the participants need to solve some hard problems before they send a message to a system, in which process the problem is hard to finish but easy to verify. Current bitcoin system is using PoW to let the miners to calculate a hash value smaller than a certain threshold value which is hard to finish but easy to verify due to the features of hash functions. However, the hash values themselves were a bit waste of calculation energy.
\fi

%----------------------------------------------------------------------------
\noindent \textbf{Proof-of-Deep-Learning (PoDL)}
In this paper, we inherit the block acceptance policy of PoDL ~\cite{chenli2019energy,li2019exploiting} to substitute the hash calculation with segmentation model training. We briefly explain the mechanism before introducing our novel protocols that addresses PoDL's limitations.

Each block interval is divided into two phases, Phase 1 and Phase 2. At the beginning of Phase 1, a task publisher releases the training dataset (with labels) as well as the hyperparameters of the deep learning model to miners and full nodes. During Phase 1, miners train their own deep learning model, and commit their model by the end of Phase 1. The commit process is completed through submitting the hash of their trained models as well as their IDs. At the beginning of Phase 2, the task publisher releases the test dataset to miners and full nodes, and each miner submits (1) the block header and the block that contains information describing the trained model on top of existing attributes, (2) the trained model, and (3) the accuracy of the trained model, to full nodes. Note that the hash of the block header does not need to be smaller than the threshold because the hard computation is replaced with the model training. Full nodes, during Phase 2, validate the  submitted models to check whether they have the claimed accuracy, and this happens on top of existing validation in the blockchain (\textit{e.g.,} validation of the correctness of transactions, Merkle tree, hash). To avoid miners over-fitting their models on the disclosed test dataset or stealing others models (published during Phase 2), full nodes discard any block whose model was not committed during Phase 1 (\textit{i.e.,} hash of the model or ID have not been received in Phase 1). Full nodes will accept the block that is submitted with the highest-accuracy model that claimed its accuracy correctly. They choose and validate the models in decreasing order of the claimed accuracy for that.
Such a block acceptance policy yields a robust consensus and is secure against double-spending attack as long as no more than 51\% of computation power is owned by the attackers. 
However, the PoDL is limited in that they can only handle one task at once, and there is insufficient details about how to handle training model.

To address these drawbacks, we present an alternative PoW mechanism that asks miners to perform biomedical image segmentation tasks and present a trained segmentation model as the proof. The major contributions of this paper are: (1) our blockchain allows submissions of multiple tasks, (2) our blockchain can handle large models with large training datasets. These contributions are significant since they make the idea of Proof-of-Useful-Work behind the PoDL more practical and applicable in the real world by supporting multiple tasks and larger predictive models, (3) our blockchain suggested a sustainable task scheduling mechanism which provides incentive for task publishers, thus the system will be sustainable. 

There exists other consensus mechanisms based on deep learning as well.
Privacy‐Preserving Blockchain Mining ~\cite{turesson2018deep} proposed a Sybil-resistance scheme based on privacy-preserving machine learning as PoUW consensus. In their design, they applied the hybrid consensus protocols ~\cite{pass2017hybrid} which dynamically selects flexible amount of full nodes as committee members and the participants include data providers, miners, the committee and non-committee nodes. In addition, Privacy-Preserving Blockchain Mining ~\cite{turesson2018deep} introduced their two parallel chains that long-interval for useful work and short-interval for transaction. In the simulation, they especially evaluated the submission conflict scenario. 
Coin.AI~\cite{baldominos2019coin} introduced a frame which requires miners to train DL models as PoUW and a proof-of-storage scheme for rewarding users.
WekaCoin~\cite{bravo2019proof} presented a new distributed consensus protocol which alleviates the computational waste in PoW brute-force algorithms and creates a public distributed and verifiable database of DL models.

All three related works address the first challenge which supports miners to perform DL training as PoUW.
Privacy‐Preserving Blockchain Mining and WekaCoin may not address the seconnd challenge where an attacker could steal and improve the pre-trained model which submitted from honest miner, and submit improved model before the block is conformed.
The Coin.AI addressed the second challenge which required miners to train deep learning models with hyperparameter achieved from the hash value of the previous block. With a pre-fixed hypermarameter, attackers will have less opportunities to improve the honest miner's DL model. 
Here, DLBC applied embedded watermark in the DL model, therefore, the honest miner can claim the ownership of a model and full nodes will detect the DL model from attackers. 

In general, all three related works described similar participants with different names. They all give sufficient incentive for miners. But, it is also important to collect tasks for miners to train and all three related works did not provide a proper strategy to encourage task publishers when the system needs more training task. In DLBC, we introduced an algorithm to provide reward for task publishers when the task queue is relative short. In addition, we introduced ranking score which considered the training difficult, data size, model size, network, and task queue. 

%Our approach is different from all related works. From aspect of structure, our innovation includes augmented block header, block data, memory pool, three phases of block interval. We also provided a solution to create incentive for task publisher. 

\noindent \textbf{Other Proof-of-Useful-Work mechanisms:}
Primecoin \cite{king2013primecoin} is an altcoin that asks the miners to find a special sequence of prime numbers (Cunningham chain) instead. Although the outcome of miners' computation has mathematical and research meaning, \textit{i.e.,} discovering the Cunningham chain. The application of Cunningham chain in the real world is unclear.
\iffalse
Primecoin \cite{king2013primecoin} is an altcoin that asks the miners to find a special sequence of prime numbers (Cunningham chain) instead. Although the outcome of miners' computation has certain meaning, \textit{i.e.,} discovering the Cunningham chain, the Cunningham chain does not have a clear application, and whether the work performed by the miners is useful in real world is not clear.
\fi

Proof of Exercise (PoX) is a design proposed in~\cite{shoker2018brief}, which is another PoUW mechanism that lets miners perform certain exercises and present the outcome as a proof. In PoX, \textit{employers} publish their tasks onto a board and the miners will randomly fetch tasks from it. The limitation of PoX is that they rely on this centralized board maintained by a third party, which significantly dilutes the decentralization property of the blockchain.
%Proof of Deep Learning (PoDL) was proposed by xxx. The main contribution of PoDL was that they substitute the hash calculation in bitcoin system by letting miners train some deep learning models. In their design, they assumed that there is only one model in each block generation process and the model they use was a very simple one.

There are similar but orthogonal approaches as well.
Hybird Mining~\cite{chatterjee2019hybrid} and Conquering Generals~\cite{loe2018conquering} described a similar mechanism which solves NP-complete problem~\cite{li2019privacy} as the proof in their PoUW consensus. Proof-of-Search~\cite{shibata2019proof} addresses the energy waste issue in PoW by solving optimization problems as PoUW.

\section{Definitions and assumptions} \label{sec:def}
%------------------------------------------------------------------
%

In this section, we define the entities involved in our blockchain, which will be used to support DL training as PoUW.

\subsection{Participants}
%definition
\noindent \textbf{Miners} 
are the machines of individual or small organizations who wish to contribute their computation power for maintaining a blockchain and may receive rewards as the exchange. In our case, we only consider a standard computer (not ASIC machine) with one or more dedicated graphic cards as a miner, for instance, the gaming machines and deep learning machines. Miners train the DL models as PoW with GPUs, maintain a max heap of submitted tasks based on task rewards and validate the result of potential block owner. 

\noindent \textbf{Full nodes} 
record all blocks and transactions, maintain a min heap of submitted tasks based on task reward, validate the submitted tasks and check the checkpoint of miners and validate the result of potential block owner. 

\noindent \textbf{Task publishers} 
release biomedical image segmentation training tasks and training data. After a training task is selected and performed by the miners, the corresponding publisher will pay certain amount of reward to the miner presenting the best image segmentation model in the form of the cryptocurrency that is maintained by the blockchain. %Task publishers' best interest is to achieve the best image segmentation model.

%will achieve the task reward once its model is the best in accuracy and be validated by others. 

%Once the training task is finished and validated, the publisher will pay the task reward via smart contract. 

%In practice, the publisher will be required to pay the reward as a deposit into a dummy address in order to be listed in the task list. \noteTaeho{we know sending to dummy address is not supported by normal cryptocurrencies. how do publishers send the reward to the winner?}

%Task publishers release biomedical image segmentation training tasks and training data. 

\subsection{Assumptions}
There are three assumptions that our design relies on. Some of them hold naturally in existing blockchains while others do not.

\noindent \textbf{Assumption 1:} We assume task publishers' best interest is to achieve the image segmentation model with the best performance. Therefore, we assume no collusion happens between miners and task publishers, because colluding with miners (\textit{e.g.,} disclosing test datasets to specific miners) will degrade the accuracy of the model only. % We assume a task publisher's goal is to achieve a well-trained image segmentation model, and colluding with miners 
However, it is true that miners are well motivated to collude with task publishers (even though task publishers are not motivated to do so) since winning miners gain block rewards. It is our future work to achieve a robust consensus mechanism that does not rely on this assumption. 

Besides, we also assume task publishers will pay the task reward honestly once their tasks are performed by the miners. However we introduce how to relax this assumption via smart contract by the end of this paper.
%For instance, a certain percentage of task reward will be dedicate to developers. \noteTaeho{developer is not predefined. Did you mean ``miners"?} Once the value is more than the block reward, there is no motivation against policy for a publisher.\noteTaeho{what is the policy?} 

\iffalse
2. Data security or privacy problem is orthogonal to the problem addressed in this paper, therefore we simply assume all training data-set are publicly accessible and the publication of training data-set does not breach individual privacy. In case the privacy becomes an important issue, existing work on training a model with encrypted data~\cite{graepel2012ml} can be included in our further work.% The data confidential issue is less related to our current work. 
\fi

\noindent \textbf{Assumption 2:} We assume the training tasks can be interrupted and stopped at any time by the miners. We make this assumption because training tasks may be complex and time consuming, but we need to guarantee certain block generation rate. The gap between the length of training time and the short block interval will be handled by allowing the miners to stop the training tasks at any time and submit the saved checkpoint as their proof of work during the block interval. Note that this assumption holds for optimization algorithms that are based on gradient descendent.
%For a selected training task, it can be finished in one block time. This assumption is acceptable since this scenario can easily extend to solve a more complex DL training task. For instance, the unfinished part of the complex task could return back to the unselected pool, thus the task could be finished later. Furthermore, if a publisher is allowed to reserve multiple blocks to train, this can be the other simple solution. 

\noindent \textbf{Assumption 3:} The full nodes' network condition is stable and reliable enough such that all full nodes have the same view on their memory pool and that miners and task publishers can access such view without significant network delay. In addition, we assume the full nodes' clocks are synchronized up to the difference of 5 seconds. These assumptions are necessary for achieving security properties in our blockchain.

%------------------------------------------------------------------
% 

%------------------------------------------------------------------------
\section{Design} \label{sec:des}
%-------------------------------------------------------------------------
\begin{figure*}
\begin{center}
 \includegraphics[width=2\columnwidth]{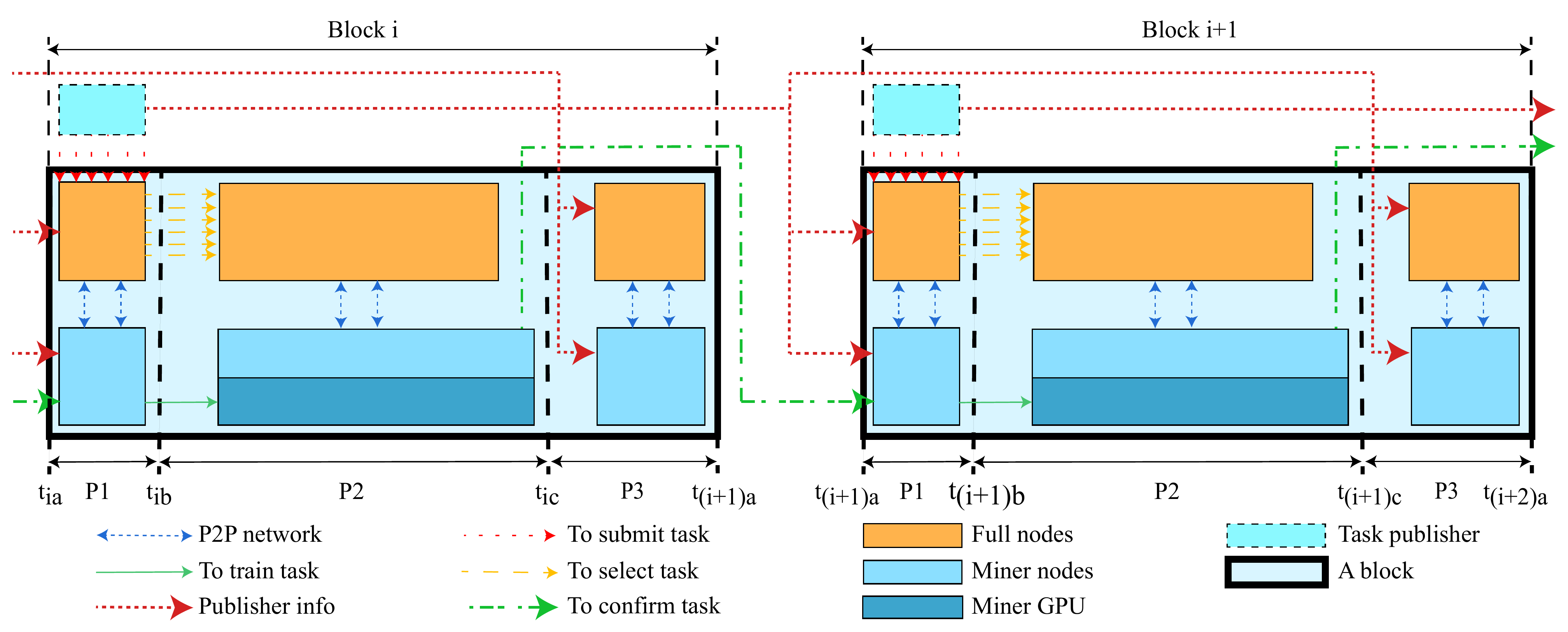}
\end{center}\vspace{-10pt}
   \caption{Description of the block mining in detail.}\vspace{-10pt}
\label{fig:scheduling}
\end{figure*}

\begin{figure}[t]
\begin{center}
 \includegraphics[width=1\columnwidth]{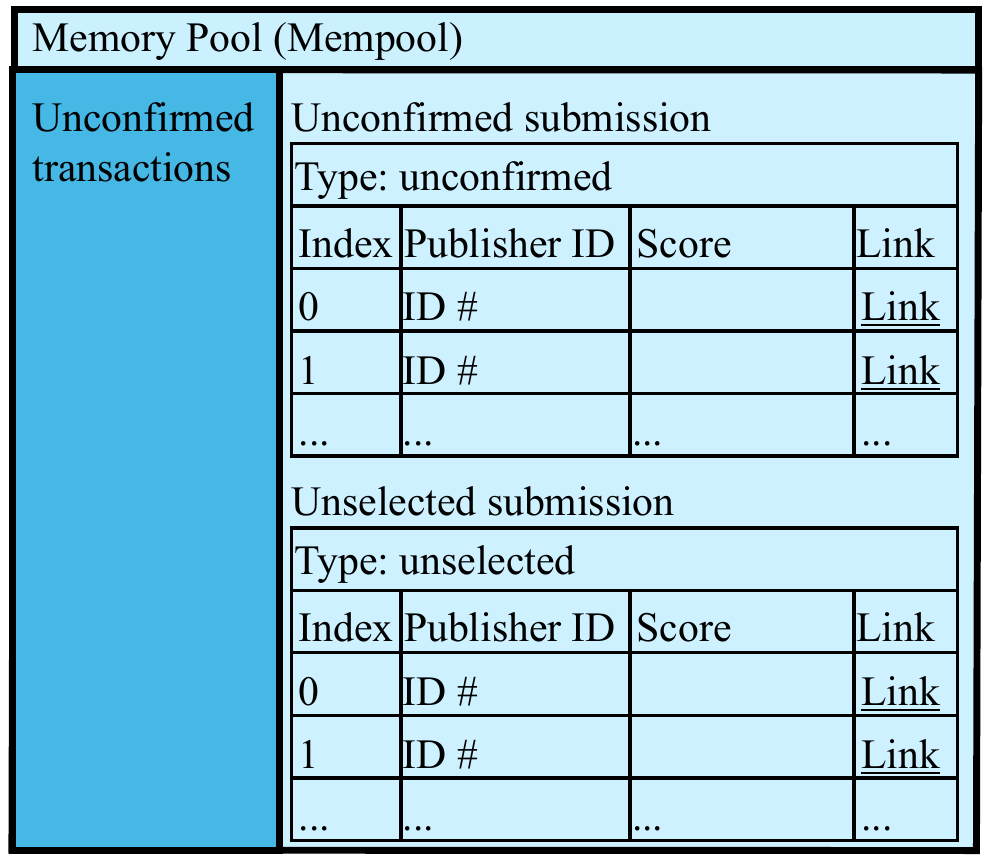}
\end{center}\vspace{-10pt}
   \caption{Our augmented memory pool. \iffalse\noteTaeho{Having the dot-lin box of the link within the mempool is confusing. Please fix it by following my email.}\fi}\vspace{-10pt}
\label{fig:mempool}
\end{figure}

\begin{figure}[t]
\begin{center}
 \includegraphics[width=.9\columnwidth]{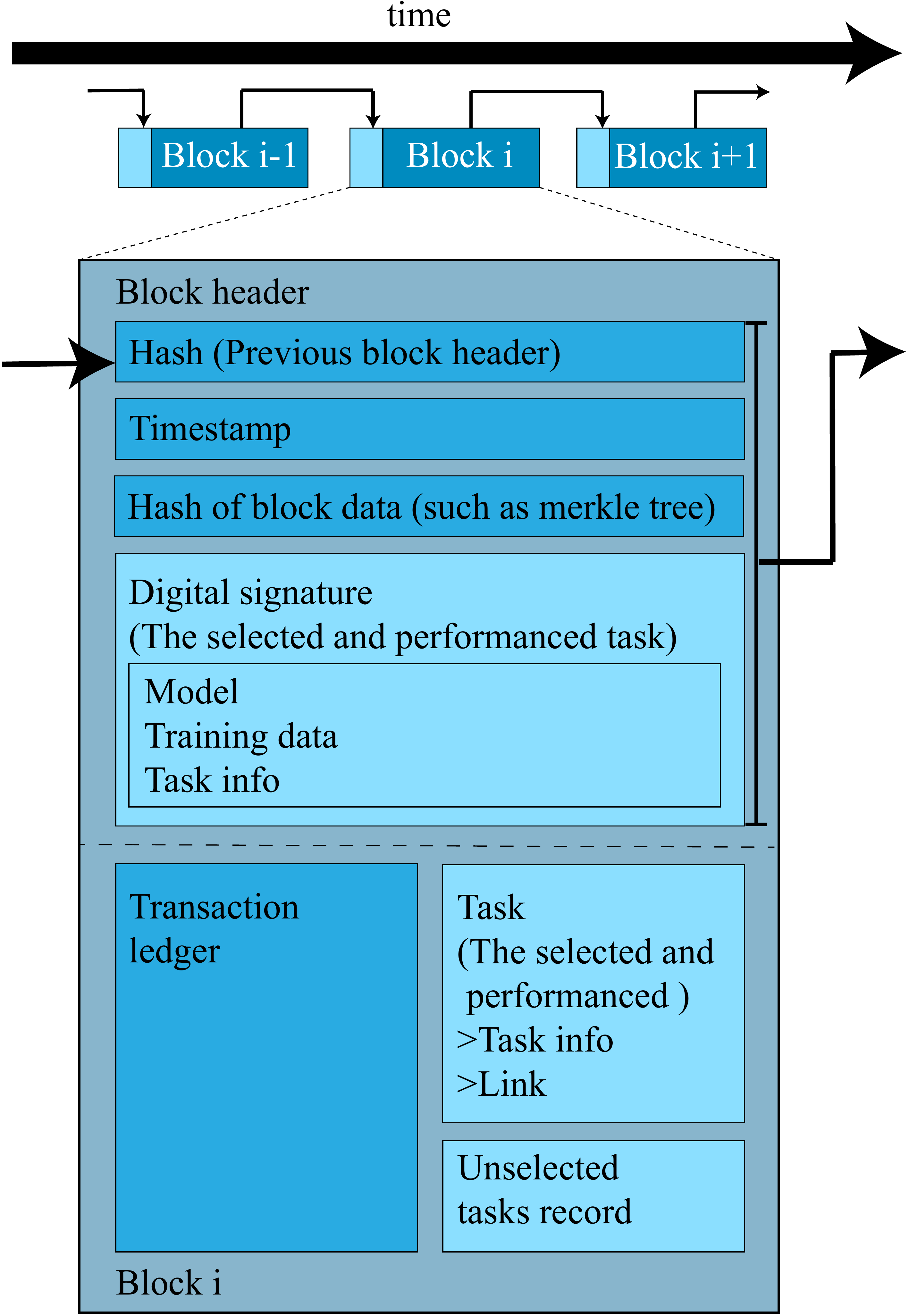}
\end{center}\vspace{-10pt}
   \caption{An overview of the block $i$.}\vspace{-10pt}
\label{fig:overview}
\end{figure}
%------------

%------------------------------------------------------------------
% general description about design and target

We propose the DLBC for maintaining a blockchain which is DL-based consensus. Most of the computation power of miners will be spent on training DL models instead of calculating useless hash values as in existing PoW mechanisms. Each new block is generated by the miner who submits the DL model with the best performance, which will be validated by the full nodes. Once the model is confirmed to be the best, the miner will generate the block and receive both of the task reward and the block reward. The block reward distribution is suggested by Equations~\ref{eq:rewardreq},\ref{eq:rewardmin}. The major novelty of this paper is to overcome the limitation of a prior work that cannot handle more than one task, large deep learning models and large training datasets. 

\subsection{Ranking Mechanism for Multiple Tasks}\label{sec:ranking-mechanism}
% As aforementioned, challenges...challenges...challenges... miners need to XYZZ.... therefore,
As aforementioned challenges about sustainability, it is important to collect sufficient training tasks from honest publishers if we can provide rewards when the task list is relative short. We introduce the equations which will schedule DL training tasks based on ranking score. 

In Equation~\ref{eq:taskq}, ${q_i}$ is the task queue index which will effect the value of ranking score and block reward distribution. The value of ${L}$ and ${l_i}$, must be positive integers, represent the estimated maximum length and current length of the task queue, respectively. Here ${k}$ is the scaling factor.  When ${k}$ is bigger than one, the threshold (${L/k}$) of the task queue length will be shorter than ${K}$. When ${k}$ is between zero and one, the threshold of the task queue will be bigger than ${K}$.

\begin{equation} \label{eq:taskq}
    q_i  = \frac{\ln{(k\times l_i)}}{\ln{L}}, \quad k>0 
\end{equation}

 In Equation~\ref{eq:diffn},~\ref{eq:diffc}, the value of ${d_{ni}}$ and ${d_{ci}}$ represent the hardness of task in terms of transmission over network and Floating-Point Operations (FLOPs) of model, respectively.

\begin{equation} \label{eq:diffn}
    d_{ni} = \frac{\textrm{model\,size} +\textrm{ data\,size}}{\textrm{median network bandwidth}}
\end{equation}

\begin{equation} \label{eq:diffc}
    d_{ci}  = \frac{\textrm{FLOPs}}{\textrm{median\,computation\,power}}
\end{equation}

As shown in Equation~\ref{eq:rank}, the ranking score of a task is based on the task reward, difficulty of the task (${d_{ni}}$ and ${d_{ci}}$) and the task queue index (${q_i}$). The Equations~\ref{eq:taskq},\ref{eq:diffn},\ref{eq:diffc},\ref{eq:rank} will suggest a ranking list of tasks based on the scores. In the current design, all miner nodes will train the task with the smallest ranking score. Thus, the system will prefer a task with high task reward, small model size and data size. When the length of the task queue is shorter than the threshold (${L/k}$), the system will prefer a complicated task. When the length of the task queue is longer than the threshold, the system will prefer a task with less FLOPs.

\begin{equation} \label{eq:rank}
    \textrm{ranking\,score} = \frac{d_{ni} + (d_{ci})^{q_i} }{\textrm{task\,reward}} 
\end{equation}

For the block reward, the Equations~\ref{eq:rewardreq},\ref{eq:rewardmin} suggest the ratio of block reward distribution that will encourage publishers to submit DL training tasks by sharing block rewards if the length of task queue is shorter than the threshold. This design will try to ensure that the task list will not be empty.

\begin{equation} \label{eq:rewardreq}
    \textrm{block}\,\textrm{reward}_{\textrm{publisher}} = max(0, 1- (\frac{k\times l_i}{L})^2) 
\end{equation}

\begin{equation} \label{eq:rewardmin}
    \textrm{block}\,\textrm{reward}_{\textrm{miner}} = 1 - \textrm{block}\,\textrm{reward}_{\textrm{publisher}} 
\end{equation}

%-------------------------------------------------------------------------
\subsection{Overview of block mining} \label{design:overview}
%-------------------------------------------------------------------------
In a traditional blockchain, the attributes of the block header, as shown in Fig~\ref{fig:overview}, include the block number, the hash value of the previous block header, the hash of the block data, a timestamp, the size of the block and the nonce value; the block data part contains a ledger that records transactions~\cite{yaga2018blockchain}. We introduce three new attributes to the block header in our blockchain: (1) digital signature of segmentation model, training data, and segmentation task information, (2) the task that is selected and performed by all miners, (3) the list of all unfinished tasks that need to be performed in the future. Each attribute will be explained in the subsequent sections.
%-------------------------------------------------------------------------

%-------------------------------------------------------------------------
In Fig.~\ref{fig:scheduling}, it shows the scheduling of tasks for block ${i}$ and block ${i+1}$. For each block, the interval is split into three partitions which named as Phase 1 (P1), Phase 2 (P2) and Phase 3 (P3), respectively.  For block ${i}$, it starts at time $t_{ia}$ and ends by $t_{(i+1)a}$. For each phase in block ${i}$, P1 starts at time $t_{ia}$ and ends by $t_{ib}$; P2 starts at time $t_{ib}$ and ends by $t_{ic}$; P3 starts at time $t_{ib}$ and ends by $t_{(i+1)a}$. The periods of P1, P2 and P3 are fixed where the length of time for P2 is much longer than those of P1 and P3.

As shown in Fig.~\ref{fig:scheduling}, the procedure will be introduced in details. In general, the length of Phase 2 is much longer than the length of Phase 3, because the training time will be significantly longer than the testing time. The Phase 1 of block $i$ starts at time $t_{ia}$. Task publishers can submit DL training tasks during this phase. The publisher will need to submit its own ID, reward, and links (model address and data address). In real world scenarios that memory pool may not hold the same view of task ranking list due to network delay, it may need an additional ledger to confirm all submitted tasks and a target task will be selected from the confirmed list. However, as it is claimed in Assumption 3, full nodes will hold the same view on memory pool, thus we will not consider this case and will address this issue in the future. In the current work, all submitted tasks will be recorded in the unselected list, as described in Section~\ref{design:Multiple}. 
\iffalse
Next, we introduce the mining process in our blockchain with details. 
At the time $t_{ia}$ (shown in Fig.~\ref{fig:scheduling}), Phase 1 for block $i$ starts, and task publishers can submit image segmentation tasks during this phase. The publisher will need to submit its own ID, reward, links (model address and data address). This task should be confirmed by the end of this block. In the worst case, this task may not be confirmed due to network delay. It will still stay in the unconfirmed tasks list for the future blocks. Once it is confirmed, the task will be moved into unselected tasks list as described above.
\fi
At the same time, the miners are training the segmentation task which was ranked at the first place by the end of the last block. After a task appeared in the unselected task list, it will join the ranking which is based on the task reward. Only the task with the highest task reward will win the chance to be trained.

Once the target training task is confirmed, miner nodes start fetching data and model, and the publisher of the selected task will release the key to the encrypted model. After the model and data are ready, each miner will evaluate the complexity of the task by training the model for one epoch. Then miners will start training with GPUs for a certain number of epochs. The number of epochs was evaluated by each individual miner and it can be different among miners. This number is measured to ensure that the miner can stop training before Phase 3, yet finish the last entire epoch. The behavior of other participators is described in Sections~\ref{design:Multiple},\ref{design:Big}.

%------------

As shown in Fig~\ref{fig:scheduling}, the primary job of Phase 3 is to test and validate the biomedical image segmentation model which was trained during Phase 1 of the current block. When the time $t_{ib}$ (shown in Fig.~\ref{fig:scheduling}) arrives, the publisher will provide API for miner nodes to test their own model and for full nodes to validate the winner model.

Miner nodes generate a digital signature which is shown as Fig~\ref{fig:overview} digital signature frame. At the same time, miners will check the accuracy of their own models. All miners will submit their accuracy values and model links. In addition, the model link is required to include all checkpoint models for verification purpose. This policy is to make sure that the final model is truly a trained model. The full nodes will sort all submitted accuracy values and verify the model with the highest accuracy. The miner, who submitted the best model, will generate the block. Meanwhile, as described in Section~\ref{design:Multiple}, the task ledger will be confirmed which also means all unconfirmed tasks are moved into unselected task list. The target task for the next block is selected from the updated unselected task list, and traditional transaction confirmation is finished.

The essential property of blockchain is that any full node should be able to verify the history data. In this case, it is necessary to check that the accepted biomedical image segmentation models are trained from training dataset only.
\iffalse
\noteTaeho{Do we require that models are trained from training dataset only? I thought we allowed randomly chosen model as well?} 
!!! If the training dataset is different, the accuracy could be different.  
\fi
In addition, the testing results must be the same as they were claimed. As described above, full nodes will be able to fetch all the checkpoint models as the reference to verify the training through the model link in the task ledger in block data.

%-------------------------------------------------------------------------
\subsection{DL model ownership protection with DNN-watermark}\label{design:watermark}
Once a miner submits a model, it is very important to protect the ownership. Otherwise, attackers will be able to steal others’ submissions to generate seemingly valid blocks without performing the actual work. For example, an attacker may attempt to steal a published model and perform short-term extra training to generate a distinct model. Such an attack will lead to a valid block since the block and the model are not bound to each other unlike the hash value and the block in PoW-based consensus.
To prevent such attacks,
we adopted DNN-watermark to protect DL model ownership \cite{uchida2017embedding}.

%%% Do not click accept without looking at it. You need to review them and learn how I write the paper.

\subsubsection{Embedding watermarks into deep neural networks}
In order to claim and protect the ownership of the deep learning model, the miner will generate a watermark and embed the watermark into the deep neural network. 

For each individual miner, they will firstly generate a unique watermark from the current block data which is derived from the hash of the block body and the block header. 
We will apply the direct watermark method as described in the DNN-watermark \cite{uchida2017embedding}. A watermark in this method is a binary matrix of $n$ columns and $m$ rows.
The method requires that each column of the matrix should contain only one 1 and the rest are 0s.  We will use the first two rows to encode the hash of the block body and the block header into the matrix, which is the watermark to be embedded into DL models. The encoding rule is to check the hash value bit by bit. If the $i$-th bit of the hash is 0, the first row of the $i$-th column is 1 in the watermark matrix; if the $i$-th bit of the hash is 1, the second row of the $i$-th column is 1 in the watermark matrix. 
Since the first transaction (i.e., CoinBase transaction) is unique, the hash value of the block data and the corresponding watermark matrix is unique unless there is a hash collision, which can be neglected since a cryptographic hash (e.g., SHA-256) is adopted in practice. %The watermark is encoded from this hash value of the block data. 

The generated watermark is embedded into DL models via regular DL training methods with an extra term in the  loss function that is optimized during the training (Eq. \ref{eq:rewardmin}). $E(w)$ is the total loss function, where $E_0(w)$ is the original loss function related to the classification errors and $E_R(w)$ is the extra regularizer term for embedding the watermark. 
%Including the original term, a watermark-related term has been introduced. 
Because a watermark-related term is introduced, the value of weights in the model will be updated to follow the requested distribution such that the watermark matrix will be embedded into the weights. This distribution of the weight values can be easily detected and this will be demonstrated in the experiment section.  As described in DNN-watermark ~\cite{uchida2017embedding}, the normal regularizer will address the over-fitting issue and this additional regularizer will train the weights of model to follow certain statistical bias. 

\begin{equation} \label{eq:rewardmin}
    E(w) = E_0(w) + \lambda \times E_R(w)
\end{equation}

\subsubsection{Determination of existence of watermark}
Once a miner submit a DL model and corresponding watermark which is unique as discussed before, full nodes will determine whether a watermark matches or not. As introduced in ~\cite{uchida2017embedding}, the watermark extraction is done by projecting DL model weights using the unique watermark from miner and it can be considered as a binary classification problem with one layer. Therefore, we will find one array of confidence level. In our experiment, we picked 0.9999 as the threshold. Only if the confidence level of all watermark elements is higher than 0.9999, the miner will be detected as honest miner and model will be confirmed as matched. Only a DL model submitted by honest miner could be considered as winner candidacy. 

% Describe how full nodes determine whether a watermark exists or not.

% i.e., above certain threshold == exists ; below certain threshold == not exists.

% Define that "we say that a watermark is embedded when XXXXX is increased beyond the threshold that a watermark is removed when XXXXX is decreased below the threshold."

\subsubsection{Ownership protection}
Note that the embedded watermark is generated from the block which is publicly known, whereas the robustness of the watermark in the original method \cite{uchida2017embedding} relies on the confidentiality of the watermark. Therefore, it is possible to remove or overwrite the embedded watermark in our scenario. For example, attackers may attempt to remove the watermark by subtracting the same extra term in the loss function instead of adding it and perform the training. 
By doing so, it is possible for the attackers to remove the watermark, but it takes many times more epochs of training to remove it as we show in the experiments (Section ~\ref{sec:exp}). It will only need less than 5 iterations of training to embed the watermark in the model, but attacker miner will need more than 30 iterations of training to remove the honest miner's watermark. Therefore, training on top of others' models will not help a malicious miner to win the block reward. In other words, by spending the same amount of computation, the attackers could achieve better models with higher accuracy if they train their own models from the scratch without needing to remove existing watermarks.

%However, removing watermark will cost additional training epochs and the model accuracy will drop significantly. The attacker will have to spend significant more time on improving the accuracy of the model. Therefore, there is additional training time for attackers as the penalty when they attempt to remove the watermark. Because of this penalty and our mechanism, the attacker will not successfully remove the watermark and claim the ownership of the block. 

\iffalse
(1) How watermark is utilized by full nodes to determine whether a model is genuine (i.e., not stolen) or not.
(2) What conditions define that a model is stolen or not.
\fi
In the third phase of each block, all full nodes will validate submissions which start from the best performance model. A full node will need to check whether the miner's watermark is matching the embedded watermark in the submitted DL model. In our experiment, only if each elements of the confidence level is higher than 0.9999, the unique miner's watermark is confirmed as matching DL model. If it matches, the model will be considered as genuine submission. If miner's watermark cannot be detected from the submitted model, the miner will be considered as malicious miner and this submission will be pruned.

\subsection{Handling multiple tasks} \label{design:Multiple}

Unlike in~\cite{chenli2019energy} where at most one task can be accepted by the blockchain, our novel blockchain is capable of accepting multiple tasks and handle them with the aforementioned ranking mechanism (Section \ref{sec:ranking-mechanism}). We achieve this by augmenting the full nodes' mempool to keep all the unselected tasks (Fig~\ref{fig:mempool}). Namely, multiple tasks submitted by publishers will reside in the mempool until they are selected and performed by the miners.

We allow task publishers to submit their tasks to full nodes only during Phase 2. To submit a task, the publisher will need to broadcast the followings to full nodes: publisher ID, task reward value, a link for downloading training dataset as well as the model (\textit{i.e.,} its hyperparameter). At the same time, the publisher will write and launch the smart contract that will send the task reward to the winning miner later when the task is performed and corresponding model is announced in the blockchain. Once the publisher submits a task, it will go to the full nodes' mempool and become an unfinished task. The unfinished tasks will stay in the mempool until the miners select and perform them. 

% be recorded in the confirmed list and will be confirmed by the end of phase two in the current block. 
%Then the task will be moved from unconfirmed tasks list into the unselected tasks list. The information on the unselected task will be fetched. If the task is selected as the highest reward task, it will be confirmed at the beginning of phase one of the next block, then miner nodes will start fetching the data and encrypted model. At the same time, the publisher of the task will release the key of the model. All other unselected tasks will be held on the list until its task reward per block is ranked as the highest.  

With multiple tasks, it becomes important to let miners agree on the same task to be performed. Otherwise, it is hard to choose the winner by choosing the highest-accuracy model, since comparison of accuracy among different tasks is meaningless. Furthermore, as we will describe in Section~\ref{section:fork}, attackers may attempt to double spend by creating forks, and it is necessary to provide a task selection for the miners to agree on one task for each block.

Our blockchain defines that all miners must choose the task with the highest reward from the unfinished tasks in full nodes' mempool (ties are broken in a pre-defined manner).
Due to the assumption that full nodes' views on the mempool are consistent, all full nodes have the same set of tasks in their mempool, and it is the blockchain policy to choose the task with the highest reward. Therefore all the miners must select and perform the same task for a specific block.

\subsection{Handling large models and training data} \label{design:Big}

In order to reduce the network traffic, a task publisher will only need to submit the model link and dataset link instead of submitting the model and the dataset directly during the task submission process. Also, to save the block storage, the link will be stored in blocks instead of actual models or datasets.

%Once the task is selected as the target, the publisher will start to release the encryption key.\noteTaeho{Readers will be confused here: ``what encryption key?'' because you did not explain why we need encryption key.} 

Miners still need to retrieve training data from the publisher, which may lead to network delay of tens of seconds or even more. To reduce this time loss, we let task publishers release the training dataset earlier in Phase 3 of the previous block's mining. After Phase 2 for block $i$, the task to be performed for block $i+1$ has been determined already, therefore the miners can start to download. Note that miners are not able to continue training in Phase 3. Otherwise, the model will be different from the one committed in Phase 1. One issue of such training data release is that the miners with high network bandwidth are advantaged because they can start mining earlier than others. %This can be utilized and lead to a new type of consensus mechanism that desires \textit{proof of bandwidth} besides proof of useful work, however we 
To avoid this and make mining fair, we let task publishers encrypt the training data with any efficient symmetric-key encryption (\textit{e.g.,} AES~\cite{mahajan2013study}) and release the encrypted training data instead. Then, the publisher releases the key at the end of Phase 3. By doing so, the network delay caused by a key is negligible (\textit{e.g.,} $\leq$ 256 bits for AES), and the miners who have finished downloading the encrypted training data can decrypt it and perform the training task immediately. The decryption causes extra delay as well, but the decryption itself can be considered as the work that miners need to prove. Note that, with the Assumption 1 in Section~\ref{sec:def}, the task publisher will not release the key to any specific miners in advance.

Symmetric-key encryption such as AES does not expand data, but it is possible that encrypted training data cannot be fully downloaded within Phase 3 because of the large volume. Motivated miners will monitor the tasks being submitted to full nodes and start fetching the training data even before Phase 3, but our mechanism may have to limit the training data to an acceptable size.

%------------%------------%------------%------------%------------%------------
%------------%------------%------------%------------%------------%------------

%------------%------------%------------%------------%------------%------------

\subsection{Handling forks}\label{section:fork}

Instead of considering the longest chain as the correct one, we let full nodes in our blockchain consider the chain who has the most highest-accuracy models as the correct one. The intuition behind this form of fork resolution is similar to that in existing cryptocurrencies based on PoW mechanisms. Namely, generating a correct block with a small-enough hash value is challenging in PoW-based cryptocurrencies, and a chain will be considered correct if it has the most correct blocks with small-enough hash values (\textit{i.e.,} being the longest chain). In our blockchain, generating a valid block with the highest-accuracy model is challenging, therefore we treat the chain with the most highest-accuracy models as the correct one. 

%\noteTaeho{I changed the fork resolution slightly. We cannot look only at  the first block causing the forks and see which chain has the highest accuracy. Instead, we need to look at all the blocks and see which chain has the most highest-accuracy models in the chain. For example, if chain A has 10 highest-accuracy model, chain B has 8, chain C has 13, then chain C is treated as the correct one. Please follow the style of Section 4 and Section 5 in http://harding.github.io/bitcoin.pdf and explain it.}

\subsection{Validating past blocks}\label{des:val}

Newly-joined full nodes need to verify the entire blockchain. When checking block $i$, full nodes will need to check the unselected tasks record from block $i-1$ and the selected task for block $i$ to see whether the task selected in block $i$ has the smallest ranking score. Here, the ranking score is given by Equation \ref{eq:rank}. The full nodes will have to verify whether the model accuracy is the same as the one claimed by the winner miner. Then, the full node will verify the digital signature we introduced in the block header to verify the integrity of data. Finally, existing validation (\textit{e.g.,} correctness of hash calculation, transaction validity) will be performed.

This work has some limitations at the current version. We assume full nodes have consistent view as well as synchronized time clock. Achieving a design with the same robustness against various attacks without relying on these assumptions is our immediate future work.
Besides, we store all unconfirmed tasks in the block, however the block size is limited to several megabytes. This limits the total number of tasks that can be handled by our blockchain. Breaking this limit is another future work.

\iffalse
\noteTaeho{Add how to verify past blocks here...}
\fi

%-------------------------------------------------------------------------
\subsection{Properties of our blockchain} \label{des:prop}

%As it was described in related literature \cite{chenli2019energy}, recycling energy from the hash algorithm of the PoW blockchain system is a novel method to share computation power. However, to maintain a reliable blockchain with the biomedical image segmentation algorithm is challenging due to larger size of the model.   \noteTaeho{this is unnecessary}

%Reduced network traffic by applying links of models

% general analysis: fixed issue and unfixed issue

%The task confirmation process in phase one will significantly decrease the network traffic during  task submission process because task publishers will only need to broadcast publisher ID 'address', ranking score and data path address, and submit task reward to the dummy deposit address instead of submitting the whole biomedical image segmentation model and training data. In this case, it is easier to synchronize the submitted tasks list between all full nodes. 

\noindent\textbf{Synchronized tasks:}
The augmented mempool stores all unselected tasks, and these mempools will be stored in every block. Miners will have to select the task with the highest reward from this list (that is available in the previous block), therefore all miners are able to agree on the task to perform. Therefore, full nodes are guaranteed to deal with the same task during the block validation. We highlight that this synchronization is achieved without relying on third-party entities, and therefore it does not harm the decentralization of blockchain.

\noindent \textbf{Redefining confirmation:} 
%In existing PoW-based blockchain, the longest chain is considered as the correct one, and the hardness of increasing the length of the chain (\textit{i.e.,} creating correct blocks) guarantees that reversing blocks is hard. Analogously, in our consensus mechanism, having more blocks containing highest-accuracy models is hard owing to the hardness of model training, and we let full nodes treat the chain with the most such blocks as the correct one in our blockchain.
Because of the way full nodes choose the next block in our blockchain, whether blocks can be reversed does not depend too much on the number of confirmations (\textit{i.e.,} the number of blocks after them on the blockchain). Rather, the accuracy of the models on the blockchain determines it. Namely, if the block contains a model with a high accuracy, it is challenging to generate another block with another model with a higher accuracy. Then, reversing the previous blocks ahead of the block with a high-accuracy model requires the amount of work needed for training a higher-accuracy model. Therefore, the blocks become hardly reversible after there being multiple high-accuracy models along the chain. Then, we may define the confirmations of a block as the number of high-accuracy models appearing after it rather than the number of blocks after it.

%become hardly reversible only after the models with high-enough accuracy appear in the blockchain. Due to this, whether blocks are reversible does not depend on the number of confirmations they receive. Rather, it depends on the highest accuracy of the model along the blocks.% Users are encouraged to treat blocks as confirmed if the accepted models' accuracy is high enough.
\noindent \textbf{Hardness of double spending:}
Full nodes will accept the blocks in Phase 3 if and only if their headers are received in Phase 1. Therefore, as long as full nodes are honest, even if adversaries delay the submission of their blocks in order to afford more time in training, they are not allowed to submit blocks with the \textit{better} models (who were trained with more time) because the block headers did not appear in Phase 1.

\begin{figure}[t]
\begin{center}
 \includegraphics[width=1\columnwidth]{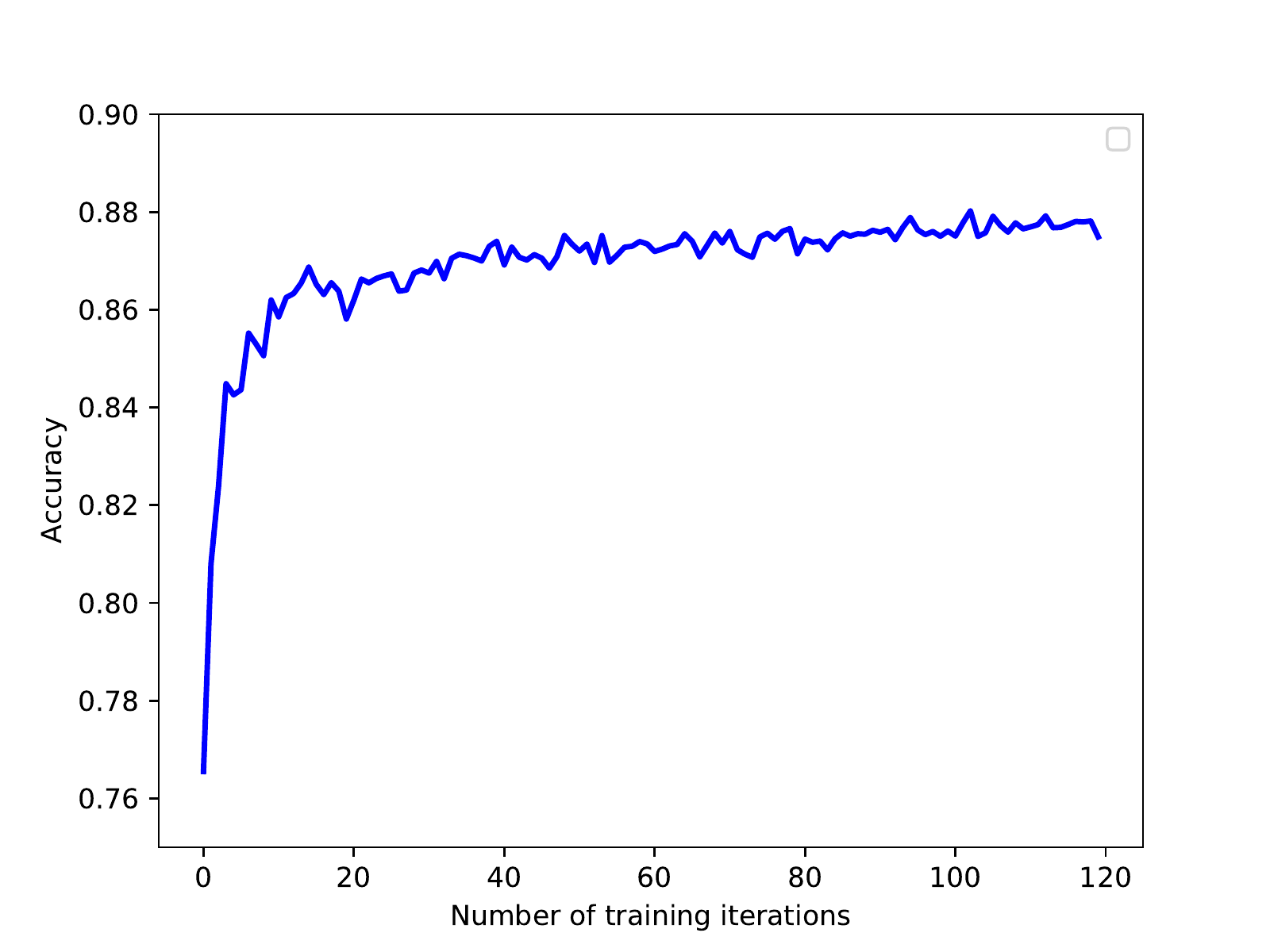}
\end{center}\vspace{-10pt}
   \caption{FCN Image segmentation accuracy results.}\vspace{-10pt}
   \label{fig:res:fcn_acc}
\end{figure}

\begin{figure}[t]
\begin{center}
 \includegraphics[width=1\columnwidth]{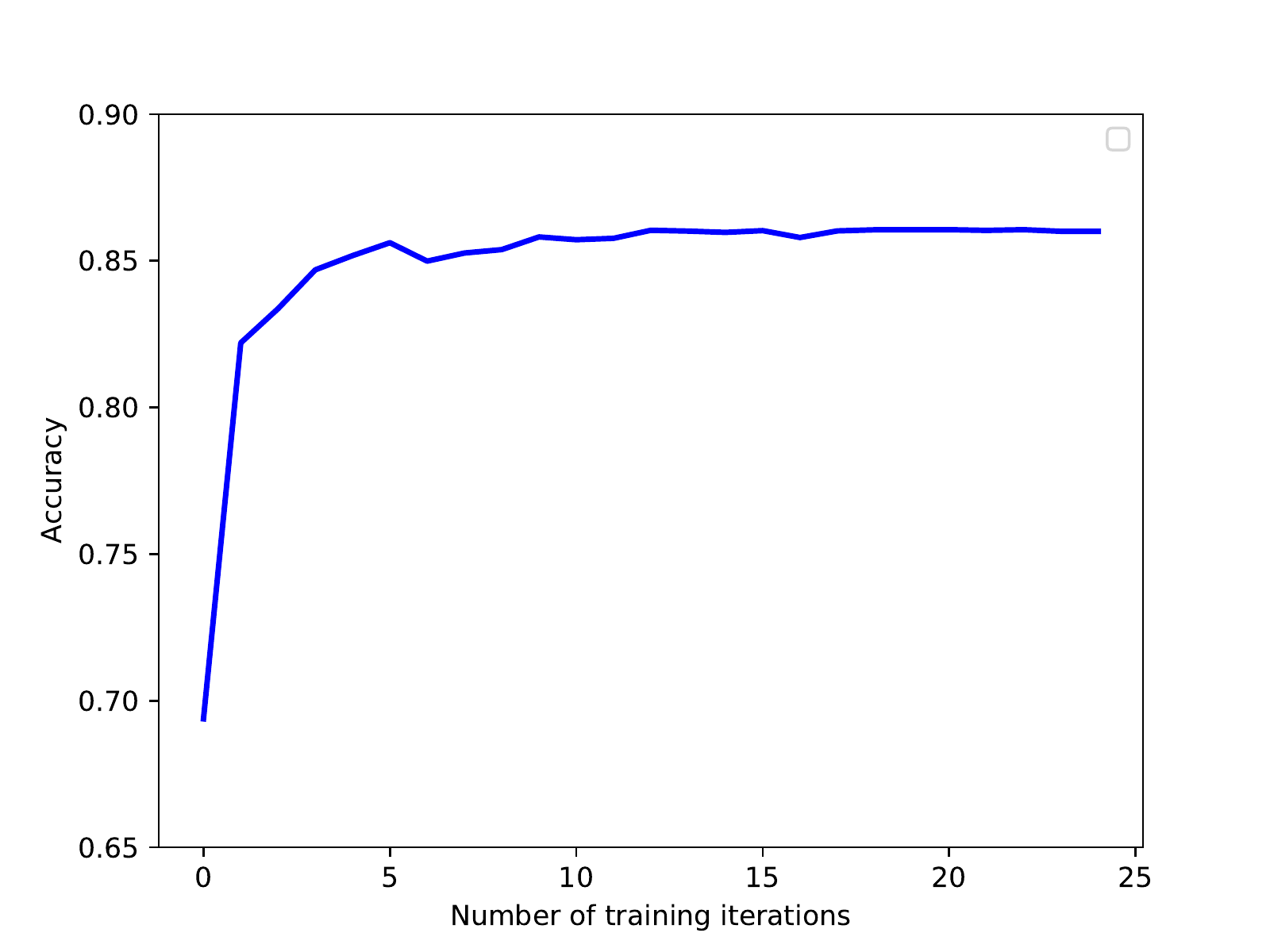}
\end{center}\vspace{-10pt}
   \caption{U-net Image segmentation accuracy results.}\vspace{-10pt}
   \label{fig:res:unet_acc}
\end{figure}

Even if the majority of the full nodes collude with miners, double spending without 51\% computing resources is still a low-probability event. During the training process, the optimization algorithms seek local optima with certain randomness because no known algorithms can strategically find the global optimum. Therefore, if only the highest-accuracy models are accepted, it is challenging to further improve the accuracy beyond it, as shown in Fig.~\ref{fig:res:fcn_acc},\ref{fig:res:unet_acc}.
\iffalse
(\noteTaeho{Can we add another figure to support this claim?}). 
\fi
If adversaries wish to double spend in our blockchain by controlling majority of the full nodes, they must present another chain with more highest-accuracy models.  %Furthermore, the biomedical image segmentation models' training must be repeatable with training dataset only as well.
Because the accuracy of the model depends on the hyperparameters and initial weights, choice of which is random, we conjecture that it is extremely difficult to generate another chain with more highest-accuracy models unless the adversary possesses more than 51\% of the computing resources for the image segmentation training. %Adversaries having good hyperparameters may have an advantage for reversing the blocks, however those parameters will be published to other miners as well, making it hard to reverse blocks again.

\noindent \textbf{Dataset and model provision:} Training dataset may have large volumes, however it is necessary for performing the published tasks. Therefore, we assume the task publishers will host training dataset for the miners' access. 

Besides, full nodes who need to verify the whole chain (\textit{e.g.,} newly-joined full nodes) need to access the historical models provided by the winning miners. We also let task publishers store the image segmentation models they collected from the miners, and provide the models to full nodes for their verification. There may be model privacy concerns, however addressing privacy concerns is an orthogonal problem, and we do not address that in this paper.

In case the storage of models (100KB-10GB per model) becomes a burden to the task publisher, we can save the storage by freeing up some earlier models with lower accuracy because the tamper-proofness is guaranteed by high-accuracy models only. Accordingly, we can also let full nodes verify the high-accuracy models only. By doing so, the blocks with high-accuracy models will still prevent the double spending, and the publishers need to store one model (the ultimate one that has the highest accuracy) per task only.

%The storage burden will be prohibitively high if dataset are stored in the blockchain, therefore we assume model requester will provision dataset properly (\textit{i.e.,} by following the release time for different blocks). Model requesters are motivated to play this role as their goal is to get the best model.
%\noindent \textbf{Storage burden:} Biomedical image segmentation models' sizes vary from 100KB to 10GB, and storing the model parameters including those for training repeatability can be a huge burden. However, various techniques can be used to reduce the sizes without affecting the accuracy too much \cite{han2015deep,gong2014compressing}, and we may limit the model sizes to a common one, \textit{e.g.,} 10MB/model in \cite{han2015deep,gong2014compressing}. 

 %In this case Considering that the model requester will accept the trained model when the accuracy does not increase significantly with new blocks, we only need to keep as many models as we need for preventing double spending (\textit{e.g.,} 5-10 blocks). 

%\noteTaeho{Need to describe that training/test dataset will be stored at model requester's side. Need to justify why model requester needs to and will be motivated to provide training/test dataset.}
%in both storage and network traffic, which is not negligible. However, considering the steady improvement in storage/network capacity, it will become manageable in the near future.
%\noteTaeho{Also need to describe the network delay of training dataset propagation along the network. It's acceptable if we keep using the same training dataset over and over.}
\noindent \textbf{Network delay:} 
Unlike the existing work~\cite{chenli2019energy},
blocks submitted to full nodes do not include the trained models any more. Instead, the block contains the links providing access to the models. Therefore blocks do not need to be very large. Our blockchain does require some extra attributes in the block header as well as various information of tasks in the block. However, the storage burden of those extra data in the block is negligible.
\iffalse
as shown in Section XXXX \noteTaeho{Analyze the extra overhead in the experiment, and cite the corresponding section of the experiment here.}, and our blockchain introduces negligible extra network delay.
\fi

However, miners' access to training data does involve non-negligible network delay which owing to the characteristics of the tasks performed by the miners. If tasks do not need to take large data as input, miners will experience less extra network delay.

\noindent \textbf{Honesty of task publishers}:
Task publishers are assumed to be honest in this paper, however this assumption can be relaxed if we adopt smart contract capable of calling external APIs. For example, if we have a function \texttt{API\_query(string URL,string link)} that sends the link of a model \texttt{link} to a web-based API \texttt{URL} and returns its accuracy against the test dataset behind the API \texttt{URL}, we can let task publishers submit and deploy a smart contract transaction 
that looks like Fig.~\ref{code:smart_contract}.
It sends the task reward (1 ether) to the message sender if s/he provides a link of well-trained model that yields a high-enough accuracy ($\geq$ 95\%) after the API call to the publisher's API for testing (\textit{e.g.,} xx.yy.com/test). Then, we can let task publishers announce their tasks by deploying smart contract transactions at the blockchain instead of announcing them to full nodes. By doing so, task publishers are unable to reject the task reward payment.
Oraclize~\cite{bertani2016understanding} can be used to implement such external API call in Ethereum-based smart contract transactions, which supports access to any API on the Internet. However, further study needs to be done to understand the security as well as burden to the full nodes, the miners, and even the task publishers.% For example, choosing the maximum-accuracy model in certain time period may be inefficient if computed in smart contract transactions, and the burden of external API calls via Oraclize is unknown either. Further study needs to be done to understand the security of such a payment mechanism.

\begin{figure}[t]
    \centering
\begin{lstlisting}[language=Solidity,numbers=none]
pragma solidity 0.4.0;
contract TaskContract {
    uint256 private reward;
    uint256 private accuracy;
    string apiForTesting;
    function TaskContract() public{
        taskReward = 1 ether;
        requiredAccuracy = 9500; // 95%
        apiForTesting = xx.yy.com/test
    }
    function testAndPay(string linkOfModel) public{
        require(API_query(apiForTesting, linkOfModel) >= accuracy);
        msg.sender.transfer(taskReward);
    }
}
\end{lstlisting}\vspace{-10pt}
    \caption{A toy example of smart contract that guarantees task reward payment.}
    \label{code:smart_contract}\vspace{-10pt}
\end{figure}

%------------------------------------------------------------------------
\section{Experiment} \label{sec:exp}
%-----------------------------------------
%---------- begin of table
\begin{table}[t]
\begin{center}
\caption{Overhead benchmark based on 1000 times testing}
\label{tab:benchmark}
\begin{tabular}{lSSSS}
    \toprule
    \multirow{2}{*}{Model} &
        \multicolumn{2}{c}{Digital signature (s)} &
        \multicolumn{2}{c}{Network (s)}\\
    & {AVG.} & {STD.} & {AVG.} & {STD.}\\
    \midrule
    {U-net (270MB)}& 1.25 & 0.051 & 3.77 & 0.322\\
    \hline
    {FCN (212MB)} & 0.98 & 0.042 & 3.01 & 0.291\\
    \hline
    {MNIST (2MB)}& 0.001 & 0.000 & 0.37 & 0.012\\
    \hline
    {ARS (3MB)} & 0.002 & 0.000 & 0.41 & 0.001\\
    \bottomrule

\end{tabular}
\end{center}\vspace{-10pt}
\end{table}
%---------- end of table

%--------------FCN res
\begin{figure}[t]
\begin{center}
 \includegraphics[width=1\columnwidth]{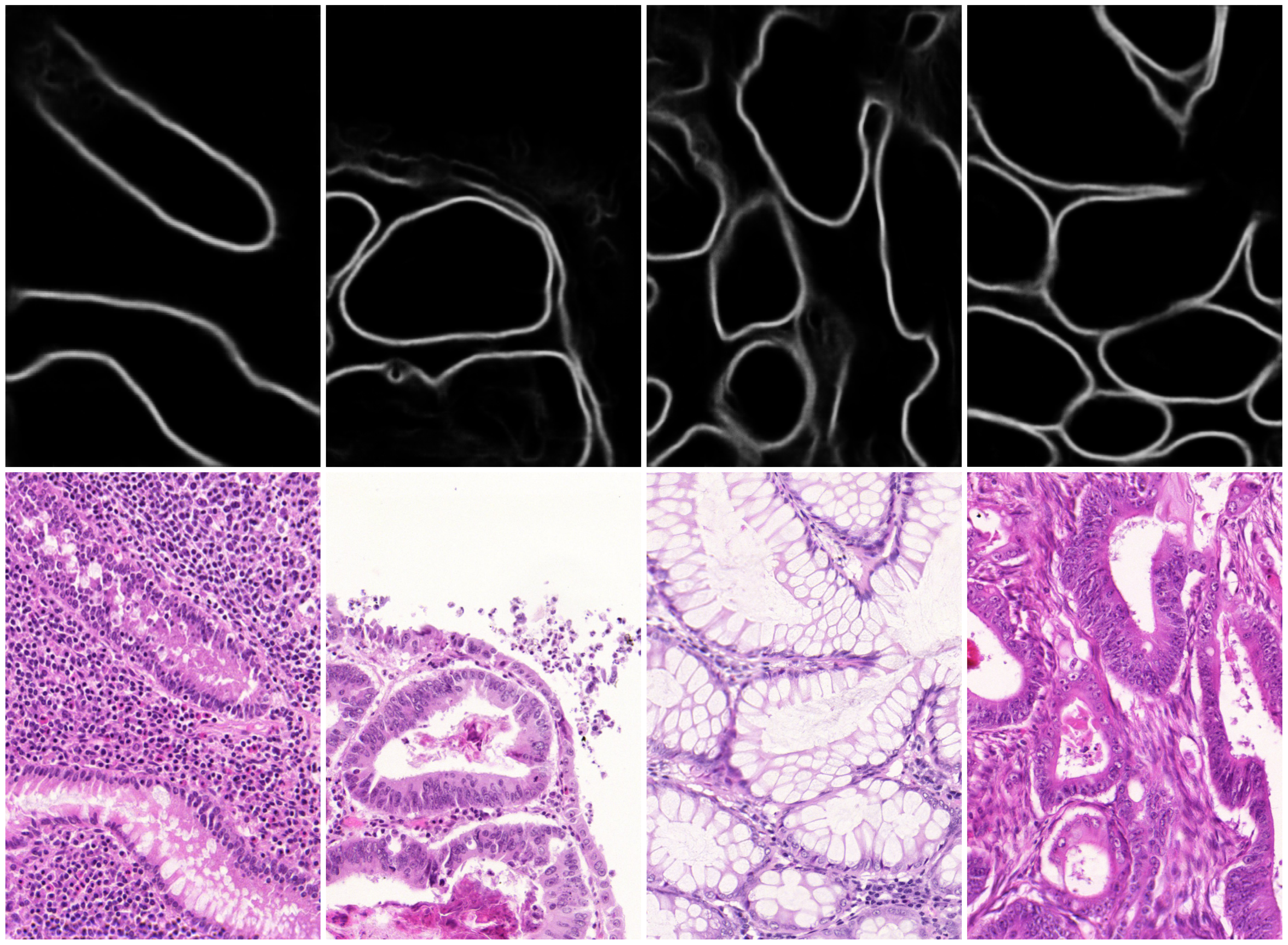}
\end{center}\vspace{-10pt}
   \caption{FCN image segmentation result. (The upper/lower row demonstrates the segmentation results and original gland histology images, respectively.)}\vspace{-10pt}
   \label{fig:res:fcn}
\end{figure}
%---------------FCN acc
%---------------Unet res
\begin{figure}[t]
\begin{center}
 \includegraphics[width=1\columnwidth]{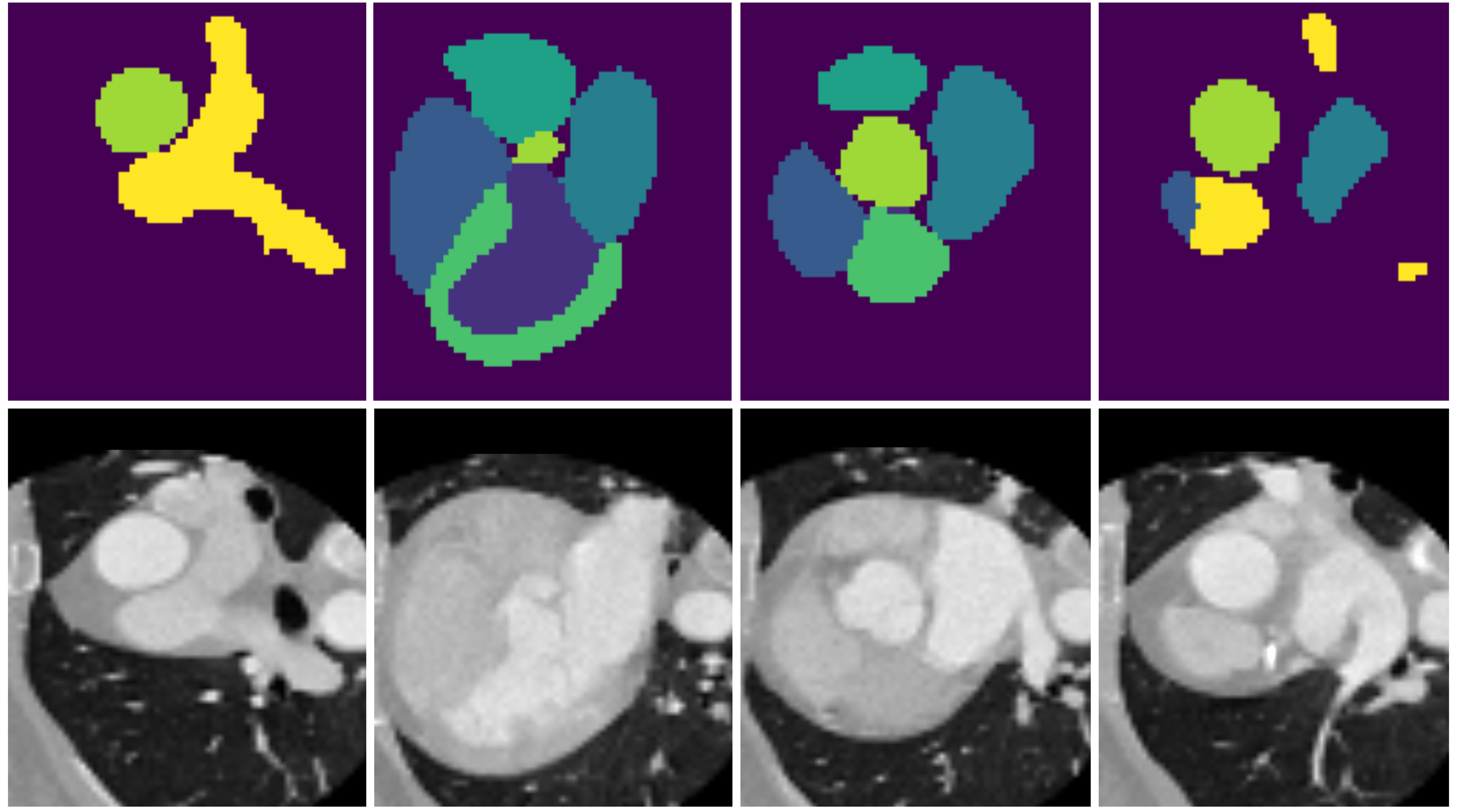}
\end{center}\vspace{-10pt}
   \caption{U-net image segmentation result. (The upper/lower row demonstrates the segmentation results and original cardiac CT images, respectively.)}\vspace{-10pt}
   \label{fig:res:unet}
\end{figure}
%---------------Unet acc
%---------------

%--------------FCN res
\begin{figure}[t]
\begin{center}
 \includegraphics[width=1\columnwidth]{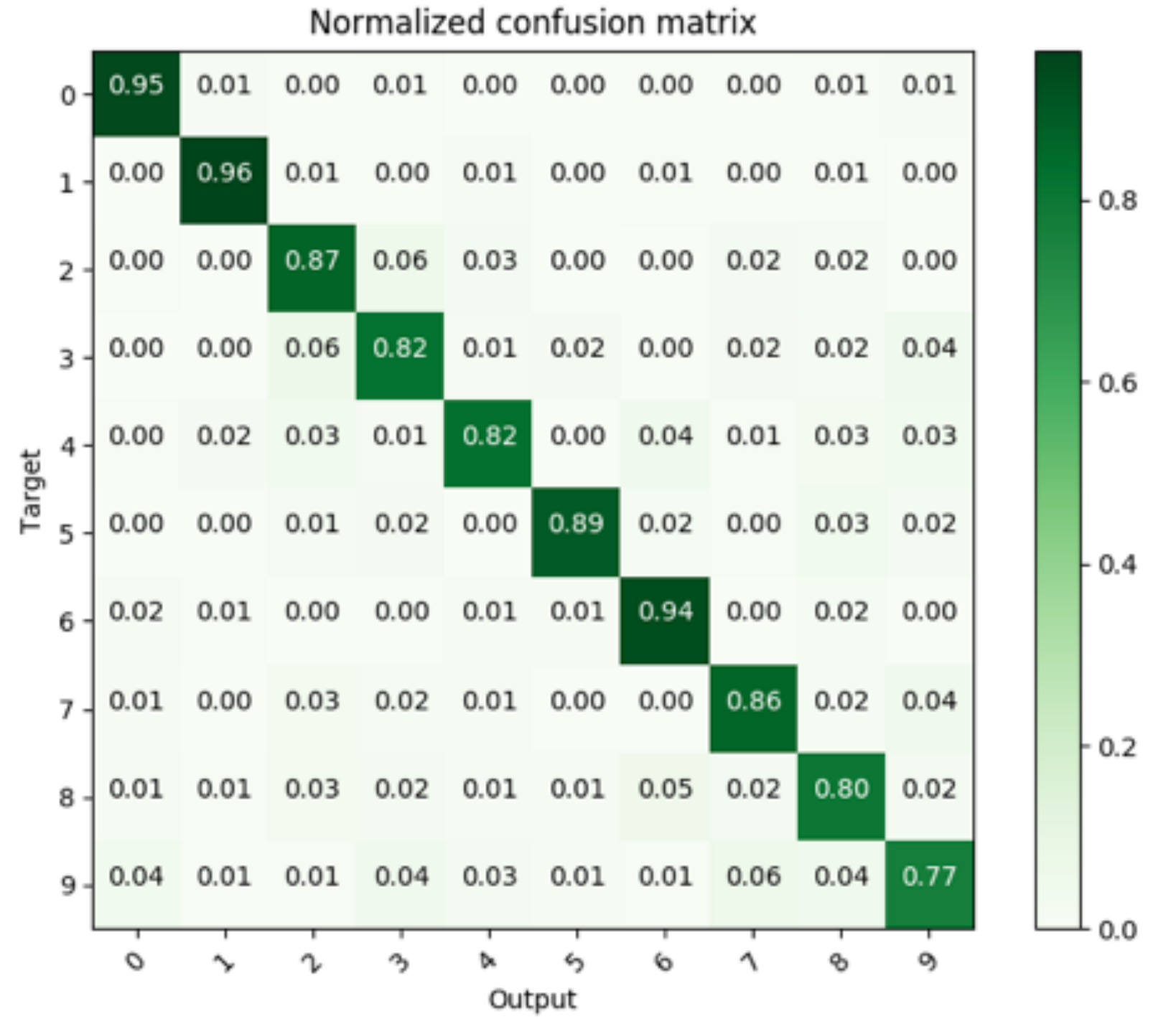}
\end{center}\vspace{-10pt}
   \caption{Confusion matrix for MNIST image classifier.}\vspace{-10pt}
   \label{fig:res:1}
\end{figure}

\subsection{Experiment setup} \label{exp:setup}

The experiments were conducted in small scale local network on the machines with Intel(R) Core(TM) i7-6850K CPU @ 3.60GHz, 32Gb RAM, GTX 1080 Ti.

To exploit computation power of blookchain for the image segmentation tasks, we adopt two widely used networks: fully conventional networks (FCN)  and U-net for for 2D and 3D biomedical image segmentation respectively.

For FCN, we adopt the same network as that in the work~\cite{yang2017suggestive}, a 34-layer FCN, which applied bottleneck design and modified the decoding part to improve the accuracy.
We use the MICCAI 2015 Gland Challenge dataset which has 85 training 2D images and 80 test 2D images.
The loss function, learning rate, regulation parameters, and training epoch are also the same as that in the work~\cite{yang2017suggestive}.
For U-net, we adopt a general configurations: (a) four resolution steps, and each resolution step contains two layers of 3 $\times$ 3 $\times$ 3 convolutions, rectified linear unit (ReLu), and 2 $\times$ 2 $\times$ 2 max pooling/up-sampling; (b) the number of filters in higher resolution step doubles that in its lower resolution step, and the initial (lowest) resolution step.
We use the CT images in MMWHS 2017 heart segmentation challenge which has 20 training 3D images and 40 test 3D images.
The loss function, learning rate, regulation parameters, and training epoch are the same as that in the work~\cite{isensee2018nnu}.
For both the two networks, we use Dice metric for evaluation.

% 3D U-net~\cite{cciccek20163d} and FCN~\cite{xu2018quantization}, as state-of-the-art image segmentation structure, are utilized as the simulation of accuracy evaluation. 
% We used existing works~\cite{xu2018quantization}~\cite{yang2017suggestive}~\cite{cciccek20163d} as the training task. On top of traditional FCN, ~\cite{yang2017suggestive} applied bottleneck design and modified decoding part of network. 3D U-net~\cite{cciccek20163d} includes analysis and synthesis path which formed of four resolution steps. For each step of the analysis path, it contains two layers of 3 x 3 x 3 convolutions, rectified linear unit (ReLu), and 2 x 2 x 2 max pooling. For each step of the synthesis path, it contains upconvolution (up-sampling), two 3 x 3 x 3 convolutions, and ReLu. There are shortcut connections between analysis and synthesis path from layers with equal resolution. 
%To evaluate the segmentation accuracy, we applied Intersection over Union (IoU) method

%--------------------------------------------------------------------------

\subsection{Benchmark tests} \label{exp:ben}
Instead of the brute-force algorithm, the miner nodes performed image segmentation tasks as described in Section~\ref{exp:setup}. Fig.~\ref{fig:res:fcn},\ref{fig:res:unet} shows the segmentation results with FCN method and U-net method, respectively. The accuracy evaluation results of FCN and U-net are demonstrated in Fig.~\ref{fig:res:fcn_acc},\ref{fig:res:unet_acc}. It can be seen that additional training based on a well performed model can hardly improve the performance of the model, thus it will be prevented from double spending as the discussion in Section~\ref{des:prop}.

Table~\ref{tab:benchmark} shows the extra overhead of digital signature and network. The digital signature was achieved by SHA-256 algorithm and the extra network overhead was evaluated by transmitting the winner model through a local network in accuracy validation step. Both overheads are much smaller than the image segmentation training time. Therefore, our mechanism utilized most of power on useful tasks and it potentially could be a contribution to both computer vision and blockchain society. Since we assumed the dataset is public accessible, the data loading time is not evaluated in the experiment. Extra storage overhead incurred by augmented mempool and novel task ledger is negligible which was discussed in Section~\ref{des:prop}.

\subsection{Ownership protection evaluation}
 Full nodes will only accept a model if the watermark of the submitted model matches the watermark of the co-responding miner. Therefore, because of the embedded watermark in the DL model, attacker will need to pay additional penalty in term of removing the embedded watermark after the attacker miner steals the winner DL models from honest miners.

\begin{figure}[t]
\begin{center}
 \includegraphics[width=1\columnwidth]{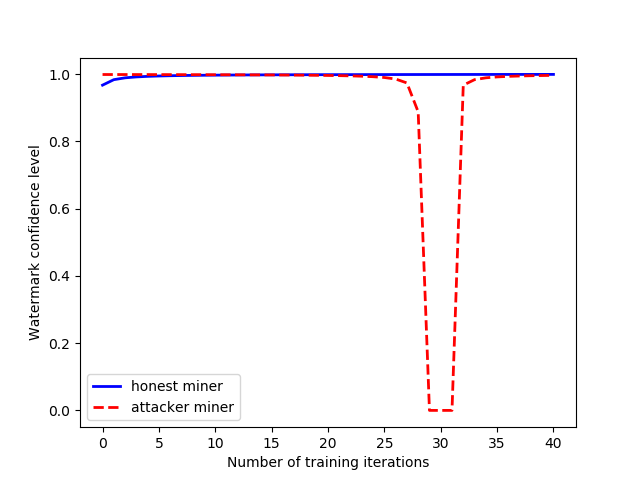}
\end{center}
   \caption{Honest miner training versus attack miner training.}
   \label{fig:res:wm:honestVSattacker}
\end{figure}

In Fig~\ref{fig:res:wm:honestVSattacker}, it shows the watermark confidence level for honest miner and attacker miner. The honest miner achieved very high watermark confidence level in less than 5 iterations. But the attacker miner will have to remove the original watermark first and it will cost around 30 iterations in our case. 
\iffalse
Because of this additional penalty, 
\fi

\iffalse
%---------- begin of table
\begin{table*}[h]
\begin{center}
\caption{Embedding watermark in 100 epochs training}
\label{tab:benchmark}
\begin{tabular}{lSSSSS}
    \toprule
    
    {lambda}& 1 & 0.1 & 0.01 & 0.001 & 0\\
    \hline
    {Accuracy (testing)}& 90.53 & 91.18 & 90.62 & 90.84 & 90.93\\
    \hline
    {Confidence} & 0.9999 & 0.9999 & 0.9995 & 0.9951 & 0.5052\\

    \bottomrule

\end{tabular}
\end{center}
\end{table*}
%---------- end of table

%---------- begin of table
\begin{table*}[h]
\begin{center}
\caption{Removing watermark in 100 epochs training}
\label{tab:benchmark}
\begin{tabular}{lSSSSS}
    \toprule
    
    {lambda}& 1 & 0.1 & 0.01 & 0.001 & 0\\
    \hline
    {Accuracy (testing)}& 90.53 & 91.18 & 90.62 & 90.84 & 90.93\\
    \hline
    {Confidence} & 0.9999 & 0.9999 & 0.9995 & 0.9951 & 0.5052\\

    \bottomrule

\end{tabular}
\end{center}
\end{table*}
%---------- end of table

%---------- begin of table
\begin{table*}[h]
\begin{center}
\caption{Attacking watermark in 100 epochs training}
\label{tab:benchmark}
\begin{tabular}{lSSSSS}
    \toprule
    
    {lambda}& 1 & 0.1 & 0.01 & 0.001 & 0\\
    \hline
    {Accuracy (testing)}& 90.53 & 91.18 & 90.62 & 90.84 & 90.93\\
    \hline
    {Confidence} & 0.9999 & 0.9999 & 0.9995 & 0.9951 & 0.5052\\

    \bottomrule

\end{tabular}
\end{center}
\end{table*}
%---------- end of table
\fi

%------------------------------------------------------------------------
\section{Conclusion} \label{sec:lim}
%---------------------------------------------------------------------------
% 

In this paper, we presented a blockchain design that lets miners to perform biomedical image segmentation model training instead of hash calculation for block mining. 
Our blockchain design addresses the limitations of existing PoUW consensus mechanisms. 
The useful work involved in our design is practical because various disease diagnosis required customized models trained on specific dataset.  
Our blockchain is able to handle multiple tasks submitted by different task publishers, and it also provides a solution to handle DNN models as well as training datasets with large size. 
We performed quantitative experiments with real-world data to show that the extra overhead introduced by our design is acceptable. 

\ifCLASSOPTIONcaptionsoff
  \newpage
\fi

% trigger a \newpage just before the given reference
% number - used to balance the columns on the last page
% adjust value as needed - may need to be readjusted if
% the document is modified later
%\IEEEtriggeratref{8}
% The "triggered" command can be changed if desired:
%\IEEEtriggercmd{\enlargethispage{-5in}}

% references section

% can use a bibliography generated by BibTeX as a .bbl file
% BibTeX documentation can be easily obtained at:
% http://mirror.ctan.org/biblio/bibtex/contrib/doc/
% The IEEEtran BibTeX style support page is at:
% http://www.michaelshell.org/tex/ieeetran/bibtex/
%\bibliographystyle{IEEEtran}
% argument is your BibTeX string definitions and bibliography database(s)
%\bibliography{IEEEabrv,../bib/paper}
%
% <OR> manually copy in the resultant .bbl file
% set second argument of \begin to the number of references
% (used to reserve space for the reference number labels box)
%\begin{thebibliography}{1}

%\bibliographystyle{ieee}
%\bibliography{bib/bib.bib}

{\small
\bibliographystyle{IEEEtran}
\bibliography{bib.bib}
}

%\end{thebibliography}

% biography section
% 
% If you have an EPS/PDF photo (graphicx package needed) extra braces are
% needed around the contents of the optional argument to biography to prevent
% the LaTeX parser from getting confused when it sees the complicated
% \includegraphics command within an optional argument. (You could create
% your own custom macro containing the \includegraphics command to make things
% simpler here.)
%\begin{IEEEbiography}[{\includegraphics[width=1in,height=1.25in,clip,keepaspectratio]{mshell}}]{Michael Shell}
% or if you just want to reserve a space for a photo:

\begin{IEEEbiography}[{\includegraphics[width=1in,height=1.25in,clip,keepaspectratio]{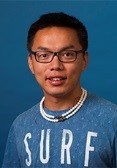}}]{Boyang Li}
Boyang Li is currently pursuing a Ph.D. degree with the Department of Computer Science and Engineering at the University of Notre Dame, Notre Dame, IN, USA. He received his B.S. degree in Electrical Science and Technology from Xi’an University of Post and Telecommunication, Xi’an, Shaanxi, China. He received his Master’s Degree in electronics and computer engineering from the University of Southampton, Southampton, Hampshire, UK. His research interests include machine learning, optimization, and novel blockchain mechanism. His paper has won the best student paper award (IEEE BIOMETRICS COUNCIL, 2019)
\end{IEEEbiography}

\begin{IEEEbiography}[{\includegraphics[width=1in,height=1.25in,clip,keepaspectratio]{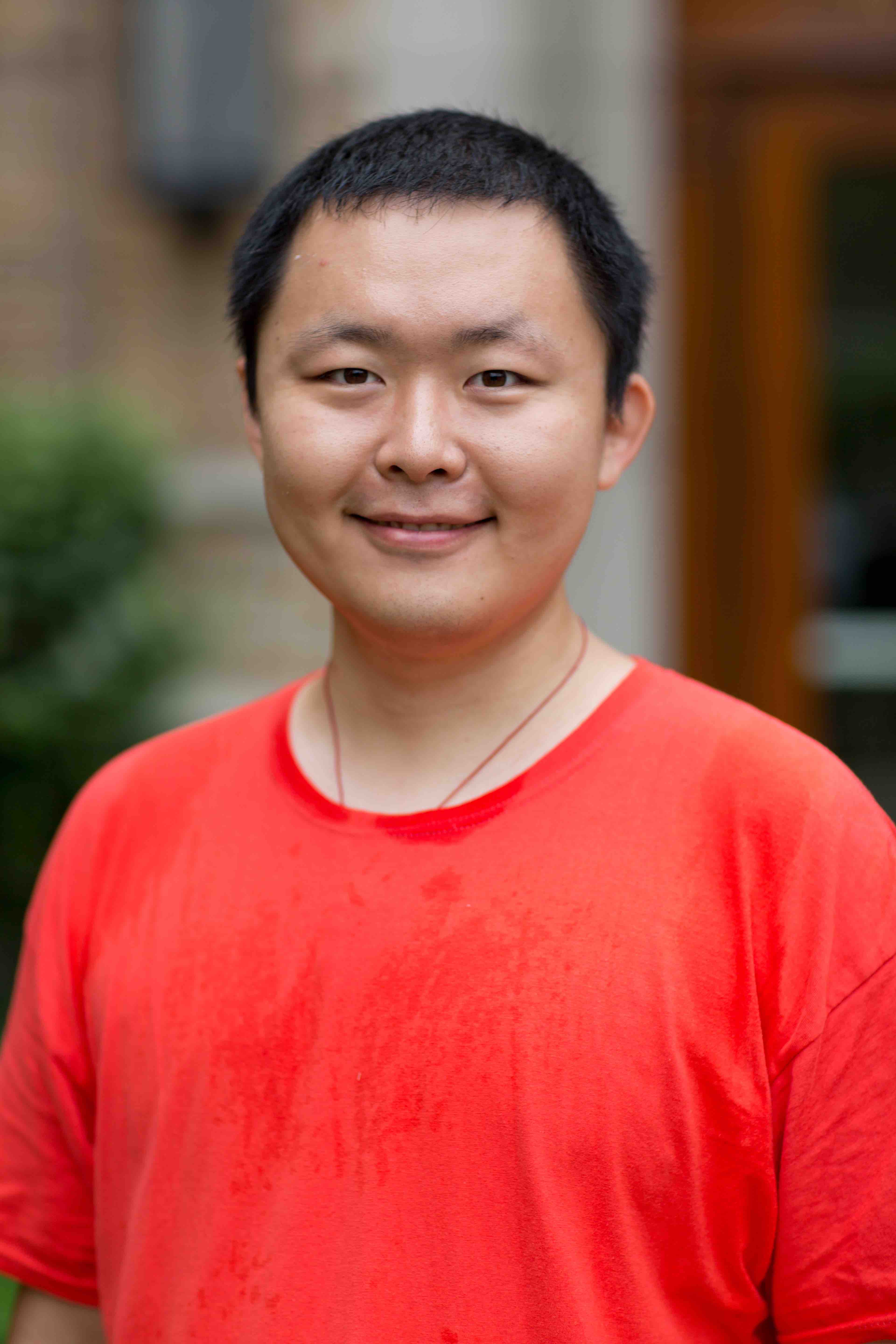}}]{Changhao Chenli}
Changhao Chenli is currently pursuing the Ph.D. degree with the Department of Computer Science and Engineering at the University of Notre Dame, Notre Dame, IN, USA. He received his B.S. degree and M.S. degree in Renmin University of China of information security and software engineering, Beijing, China, in 2016 and 2018 respectively. His research interest includes blockchain technology, smart contract and data provenance. His paper has won a best student paper award (IEEE BIOMETRICS COUNCIL, 2019).
\end{IEEEbiography}

\begin{IEEEbiography}[{\includegraphics[width=1in,height=1.25in,clip,keepaspectratio]{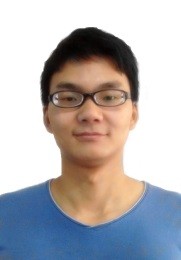}}]{Xiaowei Xu(S'14-M'17)}
 received the B.S. and Ph.D. degrees in electronic science and technology from Huazhong University of Science and Technology, Wuhan, China, in 2011 and 2016 respectively.
He worked as a post-doc researcher at University of Notre Dame, IN, USA from 2016 to 2019.
He is now a AI researcher at Guangdong Provincial People's Hospital.
His research interests include deep learning, and medical image segmentation.
He was a recipient of DAC system design contest special service recognition reward in 2018
and outstanding contribution in reviewing, Integration, the VLSI journal in 2017.
He has served as TPC members in ICCD, ICCAD, ISVLSI and ISQED.
\end{IEEEbiography}

\begin{IEEEbiography}[{\includegraphics[width=1in,height=1.25in,clip,keepaspectratio]{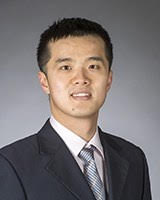}}]{Yiyu Shi(S'06-M'10-SM'15)} is currently an associate professor in the Department of Computer Science and Engineering at the University of Notre Dame, the site director of NSF I/UCRC Alternative and Sustainable Intelligent Computing, and the director of the Sustainable Computing Lab (SCL). He received his B.S. in Electronic Engineering from Tsinghua University, Beijing, China in 2005, the M.S and Ph.D. degree in Electrical Engineering from the University of California, Los Angeles in 2007 and 2009 respectively. His current research interests focus on hardware intelligence and biomedical applications. In recognition of his research, many of his papers have been nominated for the Best Paper Awards in top conferences. He was also the recipient of IBM Invention Achievement Award, Japan Society for the Promotion of Science (JSPS) Faculty Invitation Fellowship, Humboldt Research Fellowship, IEEE St. Louis Section Outstanding Educator Award, Academy of Science (St. Louis) Innovation Award, Missouri S\&T Faculty Excellence Award, NSF CAREER Award, IEEE Region 5 Outstanding Individual Achievement Award, and the Air Force Summer Faculty Fellowship. 
He has served on the technical program committee of many international conferences including DAC, ICCAD, DATE, ISPD, ASPDAC and ICCD. He is on the executive committee of ACM SIGDA, a member of IEEE CEDA Publicity Committee, deputy editor-in-chief of IEEE VLSI CAS Newsletter, and an associate editor of IEEE TCAD, IEEE Access, ACM JETC, VLSI Integration, and IEEE TCCCPS Newsletter.
\end{IEEEbiography}

\begin{IEEEbiography}[{\includegraphics[width=1in,height=1.25in,clip,keepaspectratio]{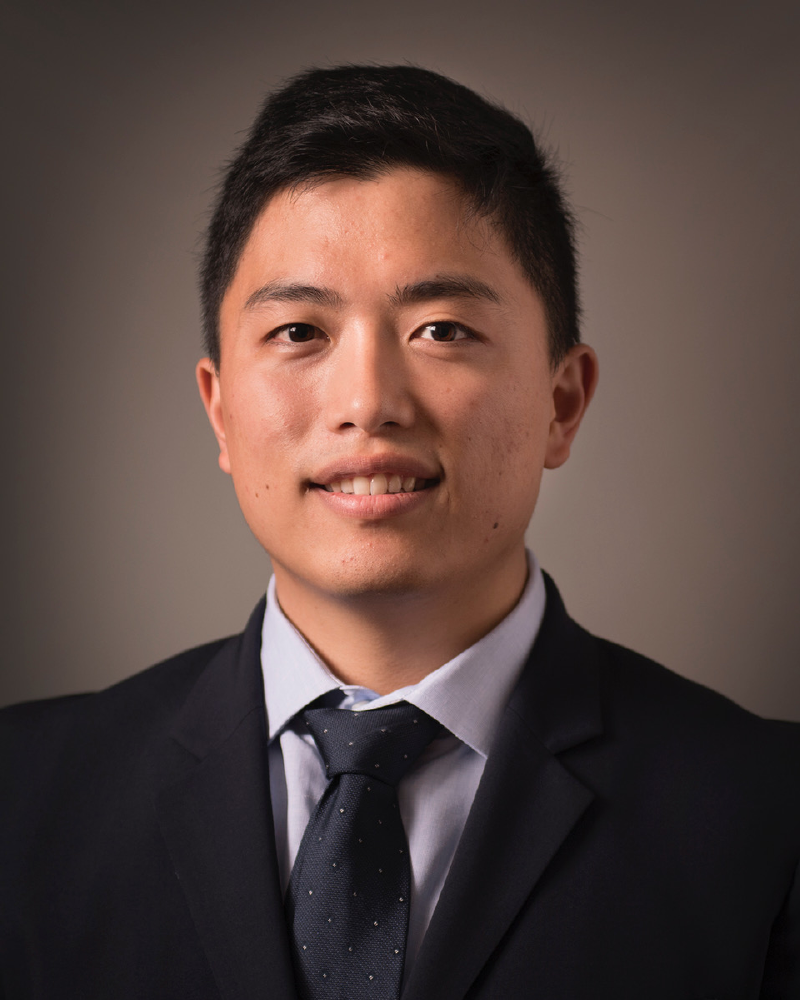}}]{Taeho Jung}
 is an assistant professor of Computer Science and Engineering at the University of Notre Dame. He received the Ph.D. from Illinois Institute of Technology in 2017 and B.E. from Tsinghua University in 2011. His research area includes data security, user privacy, and applied cryptography. His paper has won a best paper award (IEEE IPCCC 2014), and two of his papers were selected as best paper candidate (ACM MobiHoc 2014) and best paper award runner up (BigCom 2015).
\end{IEEEbiography}

\end{document}